\documentclass[twocolumn]{aastex62}
\usepackage{amsmath}

\let\oldhat\hat
\renewcommand{\hat}[1]{\oldhat{\mathbf{#1}}}

\newcommand\lsim{\mathrel{\rlap{\lower4pt\hbox{\hskip1pt$\sim$}}
    \raise1pt\hbox{$<$}}}
\newcommand\gsim{\mathrel{\rlap{\lower4pt\hbox{\hskip1pt$\sim$}}
    \raise1pt\hbox{$>$}}}

\graphicspath{{./}{figures/}}

\received{2019 November 1}
\revised{2020 January 6}
\accepted{2020 January 14}

\shorttitle{Mass Segregation in ENDs}
\shortauthors{Foote et al.}

\begin{document}

\title{Mass Segregation in Eccentric Nuclear Disks: Enhanced Tidal Disruption Event Rates for High Mass Stars}

\correspondingauthor{Hayden Foote}
\email{hayden.foote@colorado.edu}

\author[0000-0003-1183-701X]{Hayden R. Foote}
\affil{JILA and Department of Astrophysical and Planetary Sciences at CU Boulder, Boulder, CO 80309, USA}

\author[0000-0001-9261-0989]{Aleksey Generozov}
\affiliation{JILA and Department of Astrophysical and Planetary Sciences at CU Boulder, Boulder, CO 80309, USA}

\author[0000-0002-1119-5769]{Ann-Marie Madigan}
\affiliation{JILA and Department of Astrophysical and Planetary Sciences at CU Boulder, Boulder, CO 80309, USA}



\begin{abstract}
Eccentric nuclear disks (ENDs) are a type of star cluster in which the stars lie on eccentric, apsidally--aligned orbits in a disk around a central supermassive black hole (SMBH). These disks can produce a high rate of tidal disruption events (TDEs) via secular gravitational torques. Previous studies of ENDs have included stars with only one mass. 
Here, we present the first study of an eccentric nuclear disk with two stellar species. We show that ENDs show radial mass segregation consistent with previous results from other cluster types. Additionally, ENDs show vertical mass segregation by which the heavy stars sink to lower inclinations than light stars. These two effects cause heavy stars to be more susceptible to tidal disruption, which can be seen in the higher fraction of heavy stars that are disrupted compared to light stars.

\end{abstract}



\section{Introduction} \label{sec:intro}

Galactic nuclei can contain parsec scale eccentric nuclear disks (ENDs) in which stars follow apsidally-aligned eccentric orbits around a supermassive black hole (SMBH). The nearest such disk is found at the center of the Andromeda Galaxy, where HST images resolve two distinct brightness peaks, both offset from the central SMBH \citep{lauer+1993}. These peaks correspond to the collective apocenters and pericenters of stars in an END \citep{tremaine1995}. 

 ENDs are kept stable via secular, or orbit-averaged torques \citep{madigan+2018}. These secular torques drive orbits to oscillate about their equilibrium eccentricity, with orbits commonly reaching $e > 0.99$ at the peak of an oscillation, leading to a high rate of close encounters between disk stars and the central SMBH. Disk stars that pass within the SMBH tidal radius will produce tidal disruption events (TDEs) \citep{rees1988}. Additionally, \citet{madigan+2018} show that ENDs develop a negative eccentricity gradient, where stars at lower semimajor axes have higher equilibrium eccentricities. This causes stars at the inner edge of the disk to be preferentially disrupted, as low angular momentum orbits are more easily torqued to extremely high eccentricity where their pericenters dip below the tidal radius.

Several dozen candidate TDEs have been observed as nuclear optical/UV \citep{van-velzen+2011,gezari+2012,arcavi+2014,chornock+2014,holoien+2016} and soft X--ray (\citealt{auchettl+2017} and the references therein) flares that are typically observable for several years. The TDE rate from a young END can be as high as one per year\footnote{Although if disruptions really occur once per year, the SMBH may not return to quiescence between disruptions, and such events would not currently be identified as TDEs.}, orders of magnitude higher than estimates of TDE rates from an isotropic star cluster $\sim10^{-4}$ \rm{yr}$^{-1}$ \rm{gal}$^{-1}$ \citep{magorrian+1998, wang&merritt2004, stone&metzger2016}. This suggests a significant fraction of TDEs could come from ENDs. Additionally, many optical/UV selected TDEs are in post--merger E+A/K+A galaxies \citep{french+2016}. This result is consistent with the formation of ENDs during mergers \citep{hopkins&quataert2010}, as these young ENDs would produce the high rate of TDEs that we observe from these galaxies.

\begin{deluxetable*}{ c | c c c c c c }[t]
\tablecaption{Initial conditions for our N--body simulations. There are two main sets of simulations, broken down into groups that share initial conditions. For each group, we list the number of simulations, the spread (three times the standard deviation, $\sigma$) of the orbit rotation angles $\theta_a$ and $\theta_l$ (described in $\S$~\ref{sec:sims}), the number of heavy stars ($N_H$), and the resulting collisional coupling parameter ($\Delta$; see eq~\ref{eq:delta} and the surrounding discussion). Each simulation has 400 light stars. The standard deviation of the third orbit rotation angle ($\theta_j$) is three degrees. \label{tab:initcond}}

\tablecolumns{7}
\tablenum{1}
\tablewidth{0pt}
\tablehead{
\colhead{Simulation} & \colhead{Simulation} & \colhead{Number of Simulations} &
\colhead{$3\sigma_{\theta_a}$, $3 \sigma_{\theta_l}$ of Heavy Stars} &
\colhead{$3\sigma_{\theta_a}$, $3 \sigma_{\theta_l}$ of Light Stars} & \colhead{$N_H$} & \colhead{$\Delta$}\\
\colhead{Set} & \colhead{Group} & \colhead{in Group} & \colhead{[degrees]} & \colhead{[degrees]} & \colhead{} & \colhead{}
}
\startdata
& $N_H=5$ & 35 & 5 & 5 & 5 & 0.384 \\
& $N_H=10$ & 44 & " & " & 10 & 0.769 \\
& $N_H=15$ & 33 & " & " & 15 & 1.15 \\
$N_H$-vary & $N_H=20$ & 39 & " & " & 20 & 1.54 \\
& $N_H=25$ & 43 & " & " & 25 & 1.92 \\
& $N_H=30$ & 38 & " & " & 30 & 2.31 \\
& $N_H=35$ & 42 & " & " & 35 & 2.69 \\
& $N_H=40$ & 39 & " & " & 40 & 3.08 \\
\hline
 & $i_H=5$ & 10 & 5 & 5 & 25 & 1.92 \\
 & $i_H=10$ & 10 & 10 & " & 25 & 1.92 \\
$i_H$-vary & $i_H=15$ & 10 & 15 & " & 25 & 1.92 \\
 & $i_H=20$ & 10 & 20 & " & 25 & 1.92 \\
 & $i_H=25$ & 10 & 25 & " & 25 & 1.92 \\
 & control & 10 & - & " & 0 & - \\
\enddata

\end{deluxetable*}

Previous studies have simulated ENDS with a single mass stellar population \citep{madigan+2018, wernke&madigan2019}. 
Here, we use N-body simulations to explore END dynamics with a stellar mass spectrum. We expect mass segregation to proceed as it does in an isotropic cluster, with heavier stars sinking inwards and light stars scattering outwards (e.g. \citealt{spitzer1987}; \citealt{bahcall&wolf1977}; \citealt{alexander&hopman2009}, hereafter AH09; see also the review by \citet{alexander2017} and the references therein). Since END stars at smaller semimajor axes have larger equilibrium eccentricities, they are more easily torqued to orbits that endanger them to tidal disruption \citep{madigan+2018}. Additionally, as in axisymmetric disks (see \citealt{alexander+2007, mikhaloff&perets2017}), we expect \textit{vertical} mass segregation to occur such that heavy stars sink to lower inclinations than light stars. 

This paper is organized as follows: In $\S$~\ref{sec:sims}, we present our methods, including setup and initial conditions of our simulations. In $\S$~\ref{sec:results}, we show the results of our simulations, including radial mass segregation, vertical mass segregation, and how these affect TDE rates for ENDs. In $\S$~\ref{sec:discussion}, we discuss how this work may be applied to real systems, and a few implications of our results. In $\S$~\ref{sec:conclusion}, we summarize our findings.

\section{Simulations} \label{sec:sims}

We use the N--Body code \texttt{REBOUND} \citep{rein&liu2012} with the IAS15 adaptive-timestep integrator \citep{rein&spiegel2015} to simulate an eccentric nuclear disk. Following the example of AH09, we use two species of stars to study mass segregation in ENDs, where each heavy star is ten times as massive as a light star, $M_H = 10~M_L$.

We vary two quantities in our simulations: (i) the number of heavy stars, $N_H$, and (ii) their initial inclination distribution, $i_H$. We refer to simulations where the former (latter) is varied as ``$N_H$-vary'' (``$i_H$-vary''). The ``$N_H$-vary'' set aims to explore how changing the strength of the mass segregation affects the dynamics of the disk and the TDE rate of each population of stars. The ``$i_H$-vary'' set studies how vertical mass segregation is affected when the orbits of heavy stars are initially inclined above the plane of the disk. 

All simulations are run for 500 orbital periods of the inner disk. For better statistics we run $\sim$40 simulations for each set of parameters.\footnote{The precise number of simulations varies as they sometimes stall (due to formation of binary systems).} We summarize the simulation parameters in Table~\ref{tab:initcond}. 

In our simulations, we set the semimajor axis of the innermost orbit, the SMBH mass, and the gravitational constant to 1, such that the orbital period at the inner edge of the disk is $2 \pi$. 
All stars initially have semimajor axes $a \in [1, 2]$ with a surface density profile $\Sigma \propto a^{-2}$, and eccentricities of 0.7. The total disk mass is one percent of the SMBH mass, and there are 400 light stars.

We initialize orbits with aligned eccentricity and angular momentum vectors. We then introduce a small scatter in these vectors via three rotations. We draw three random angles ($\theta_a$, $\theta_l$, and $\theta_j$) from a normal distribution with a standard deviation of a few degrees (see Table~\ref{tab:initcond} for details). We then rotate the angular momentum vector about the orbit's major axis by $\theta_a$, the angular momentum vector about the latus rectum by $\theta_l$, and the eccentricity vector about the angular momentum vector by $\theta_j$.
After initializing the orbits, we search for binary systems, and remove one of their stars in order to increase integration speed. The number of stars removed is of order 5 in each simulation. 

To detect TDEs, we use \texttt{REBOUND}'s built-in collision detection capability. To set a tidal radius we have to set an overall length scale for the simulations; we choose the inner edge of the disk to be at 0.05 pc. Then the SMBH is given a radius equal to the tidal radius of a $1 M_{\odot}$ star around a $10^7 M_\odot$ black hole. 
If \texttt{REBOUND} detects that a star has come within the tidal radius of the SMBH particle, we record a TDE. The stars that disrupt are not removed from the simulation and are allowed to continue on their orbits, however if they disrupt more than once, we do not count the subsequent disruption(s) in our analysis. We keep disrupted stars to simplify analysis and to keep the disk potential as constant as possible. 

For simplicity, we take the tidal radius to be the same for light and heavy stars in our initial analysis. We discuss how the tidal radius would vary as a function of stellar properties in $\S$ ~\ref{subsec:rt} and the effect this would have on TDE rates. 



\begin{figure*}
    \centering
    \includegraphics{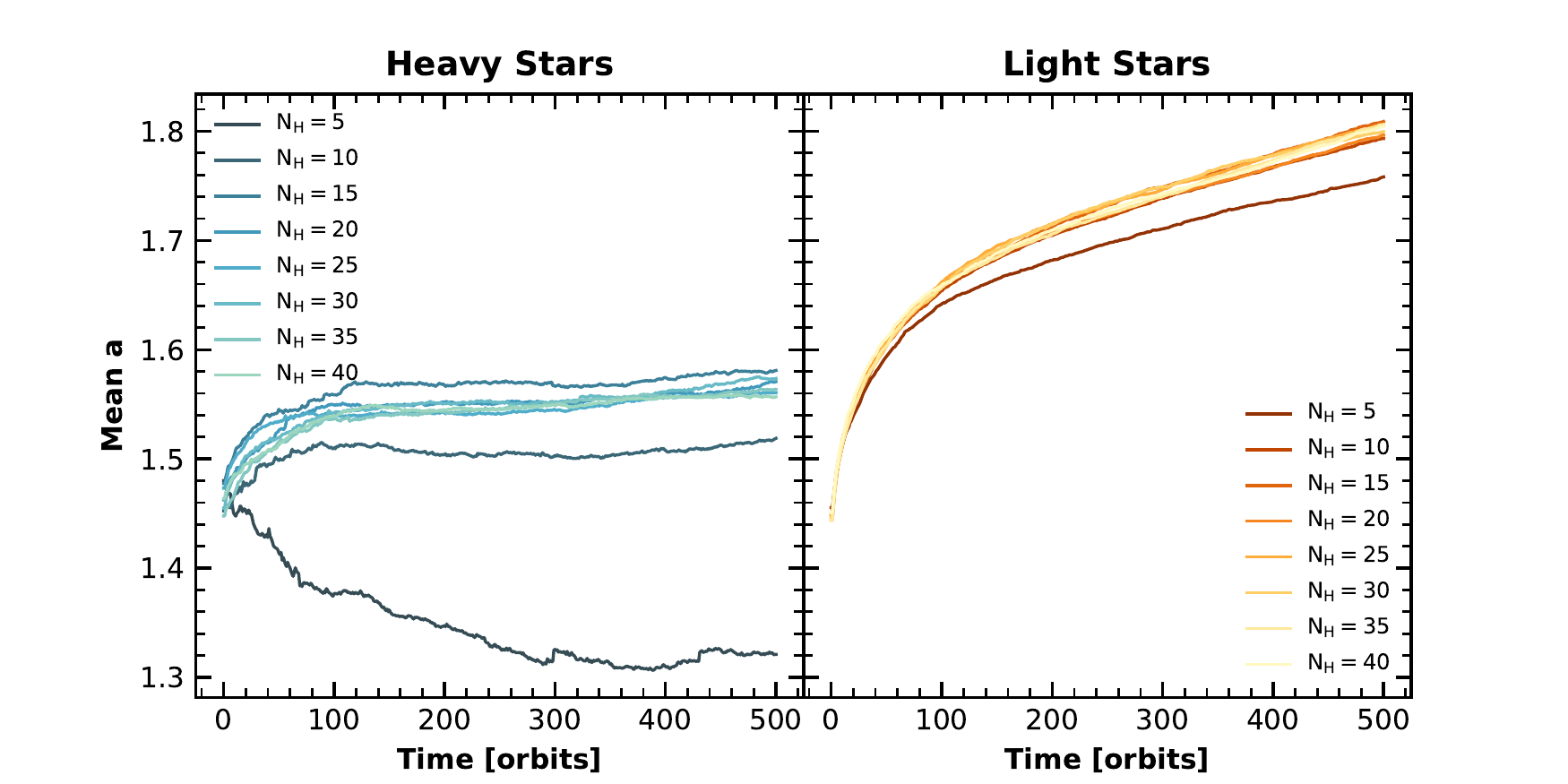}
    \caption{\emph{Left panel:} Mean semimajor axis of the heavy and light populations as a function of time for simulations with different numbers of heavy stars (``$N_H$-vary'' from Table \ref{tab:initcond}). Each line is the mean semimajor axis of all stars from all of the simulations in the group. Simulations with five heavy stars are clearly separated from the rest, suggesting the presence of weak and strong mass segregation regimes as in isotropic nuclear star clusters. In the weak regime heavy stars primarily interact with each other, while in the strong regime the heavy stars mostly interact with the light stars and sink to the inner edge of the disk via dynamical friction.
    \emph{Right panel:} Mean semimajor axis of the light population as a function of time for the same simulations.
    }
    
    \label{fig:nDistro_sma}
\end{figure*}

\section{Results \label{sec:results}}

\subsection{Radial Mass Segregation \label{rMseg}}

\begin{figure*}
    \centering
    \includegraphics{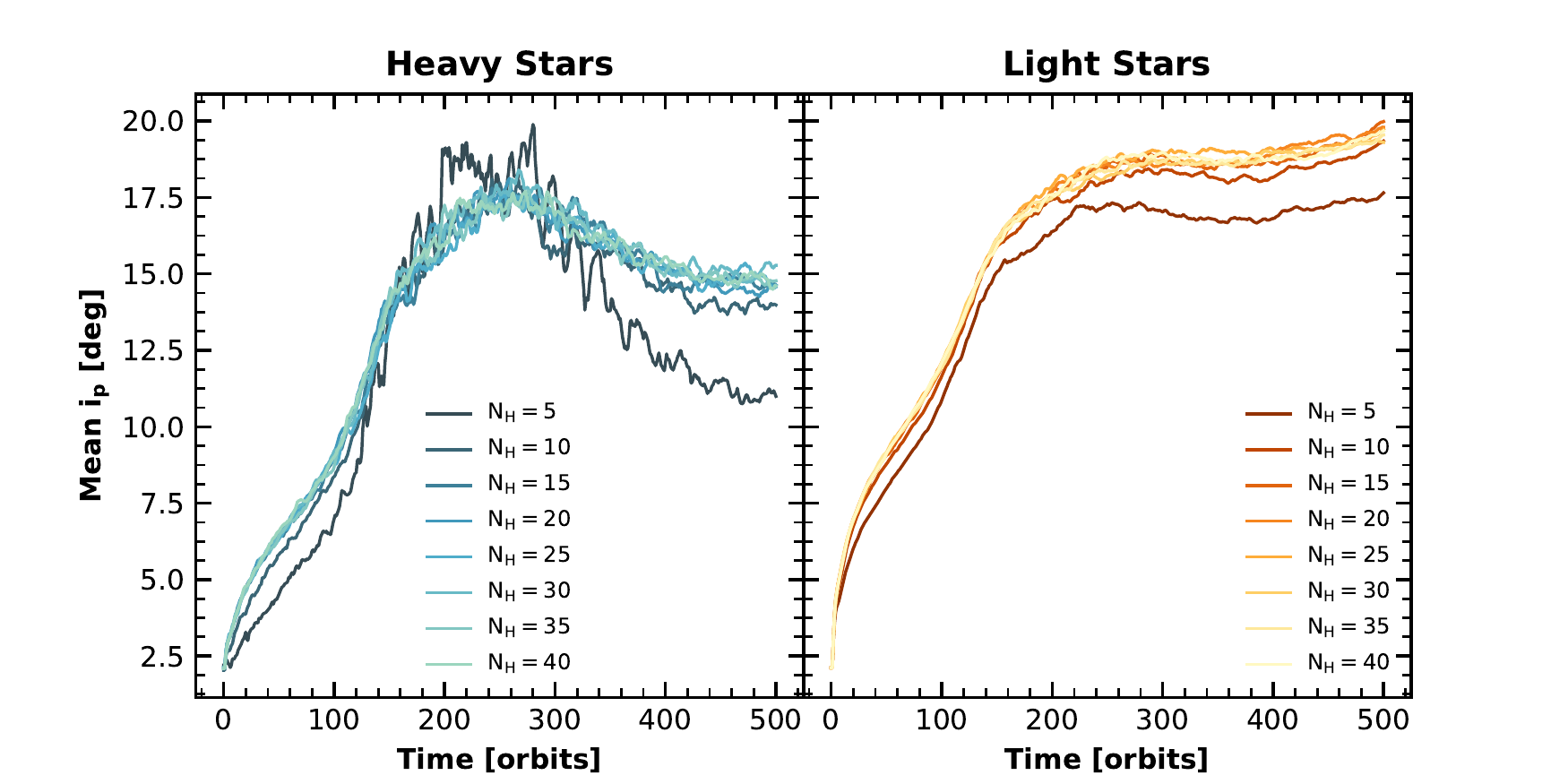}
    \caption{Mean out-of-plane inclination of the heavy and light populations as a function of time for simulations with different numbers of heavy stars (``$N_H$-vary'' from Table \ref{tab:initcond}). Each line is the mean out-of-plane inclination of all stars from all of the simulations in the group. Vertical mass segregation occurs after two secular times ($\sim200$ orbital periods). As with radial segregation, vertical segregation shows both strong and weak regimes.}
    \label{fig:nDistro_inc}
\end{figure*}

Mass segregation is the dynamical process by which massive stars sink to the center of a cluster.
Two--body interactions cause the specific kinetic energy of heavy stars to decrease while the specific kinetic energy of the light stars increases, resulting in heavy stars sinking to low semimajor axis orbits while scattering the light stars to high semimajor axis orbits. Mass segregation is well-understood in spherical clusters (e.g. AH09; \citealt{bahcall&wolf1977}; \citealt{spitzer1987}) and in axisymmetric disks \citep{alexander+2007, mikhaloff&perets2017}, but this is the first time that mass segregation has been explored in an eccentric disk with apsidally aligned orbits. AH09 studied mass segregation in an spherically--symmetric, isotropic star cluster around an SMBH using Fokker-Planck methods. They used two species of stars: heavy stars and light stars, with each heavy star having ten times the mass of a light star. This model is an approximation of an evolved stellar population, where light stars represent old low-mass main-sequence dwarfs, white dwarfs, and neutron stars with masses of order $\sim M_\odot$ and heavy stars represent stellar-mass black holes with masses of order $\sim 10M_\odot$. AH09 find that the strength of mass segregation can be parameterized by the coupling parameter

\begin{equation}
 \Delta \simeq \frac{N_H M_H^2}{N_L M_L^2} \times \frac{4}{3 + M_H / M_L},
 \label{eq:delta}
\end{equation}
where $M_H$ and $M_L$ are the mass of a heavy and light star, respectively, and $N_H$ and $N_L$ are the number of heavy and light stars respectively. $\Delta$ is the ratio of the energy--space diffusion coefficients from heavy--heavy interactions to the diffusion coefficients from heavy--light interactions.

AH09 also find that mass segregation separates into two strength regimes. In the weak regime ($\Delta \gg 1$), heavy stars are relatively common and interact with both light and heavy stars. In the strong regime ($\Delta \ll 1$), heavy stars are too rare to scatter each other frequently, and so sink to the center of the cluster primarily through dynamical friction with the larger number of light stars. As $\Delta$ decreases the heavies become more centrally concentrated, though the heavy star density profile is always steeper than the light star density profile. 

As in isotropic clusters and axisymmetric disks, ENDs show both strong and weak radial\footnote{Hereafter, we refer to mass segregation with respect to semimajor axis/energy as \textit{radial} mass segregation to differentiate it from \textit{vertical} mass segregation in a disk, which affects inclination/angular momentum.} mass segregation. Figure \ref{fig:nDistro_sma} shows the semimajor axis evolution for simulations with different numbers of heavy stars (``$N_H$-vary'' in Table~\ref{tab:initcond}). Simulations with five heavy stars are in the strong regime, where massive stars actively sink to the center of the cluster through dynamical friction. All other simulations are in the weak regime, where the heavy stars simply relax to lower semimajor axes than the light stars. The $\Delta$ cutoff between the strong and weak regime in ENDs is thus likely between 0.4 and 0.8, consistent with previous results from isotropic clusters ($\Delta \approx 1$). 

While the mean semimajor axis of all stars tends to increase as seen in Figure \ref{fig:nDistro_sma}, our simulations do conserve energy. The energy of an orbit $E \propto 1/a$. 
If a star's semi--major axis doubles, another star's semi--major axis would only have to decrease to two--thirds of its original value to conserve energy. In this example the mean semimajor axis would increase.

\begin{figure*}
    \centering
    \includegraphics{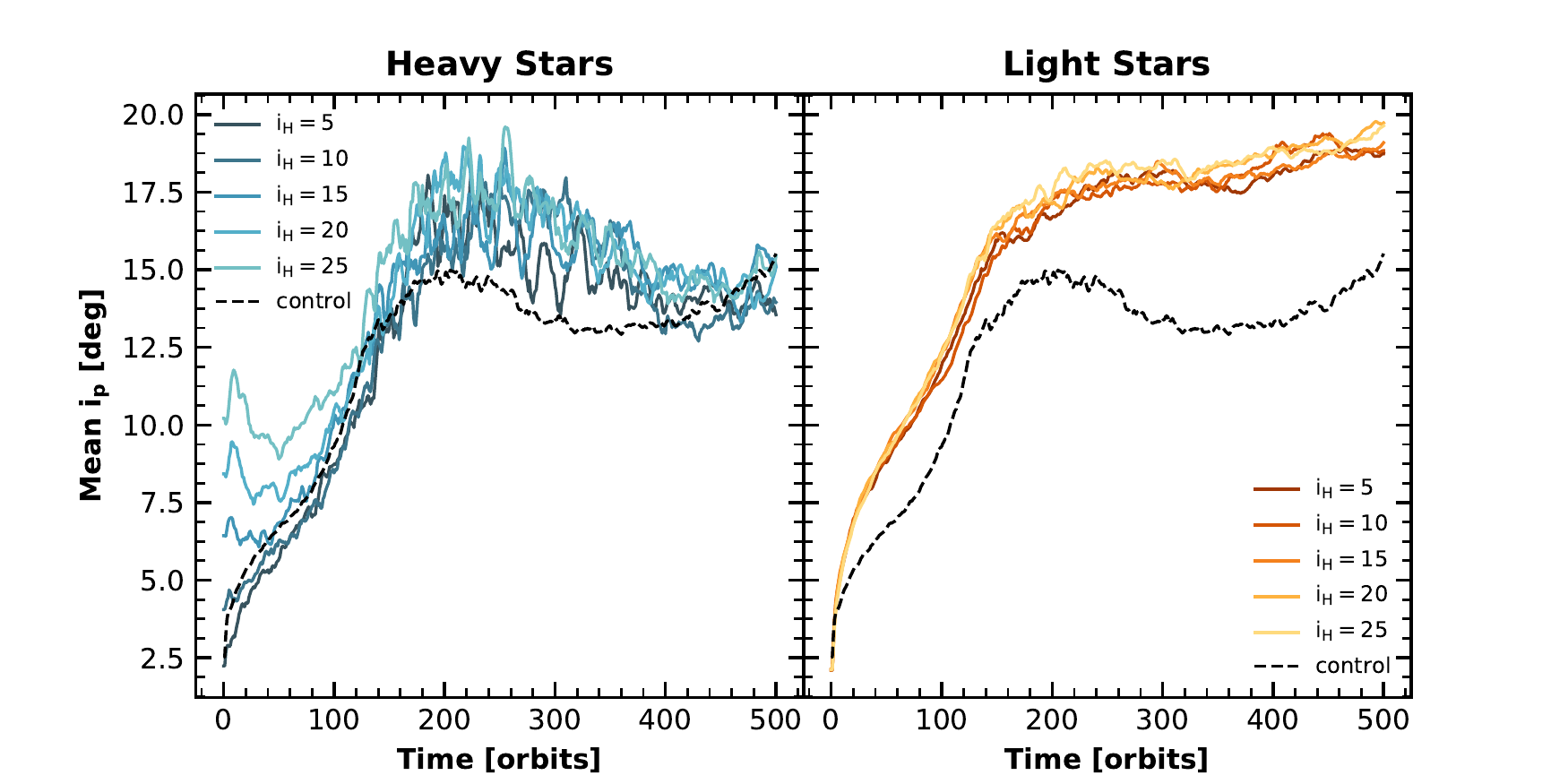}
    \caption{Mean out-of plane inclination versus time for simulations with different initial heavy star inclinations (the ``$i_H$-vary'' simulations in Table~\ref{tab:initcond}). Each solid line is the mean out-of-plane inclination of all stars from all of the simulations in the group. The dashed line shows the mean out-of-plane inclination in simulations with no heavy stars. \textit{Left Panel}: The heavy stars relax to roughly the same inclination after $\sim150$ orbital periods, regardless of their initial inclinations. \textit{Right Panel}: The light stars do not show any dependence on the initial conditions. 
    }
    \label{fig:incDistro_incBySim}
\end{figure*}

\subsection{Vertical Mass Segregation \label{vMseg}}

Vertical mass segregation is a similar process to radial mass segregation, by which heavy stars sink to low inclinations while scattering light stars to high inclinations. It has been studied in the context of axisymmetric stellar disks around SMBHs \citep{alexander+2007, mikhaloff&perets2017}. They showed that in such systems, the scale height of the disk decreases with stellar mass due to two--body relaxation.

Vertical mass segregation can also occur due to secular effects. In particular, disk orbits can perturb each others' angular momenta via orbit--averaged torques on a secular dynamical timescale, viz.
\begin{equation}
    t_{\rm sec} \equiv \left(\frac{M_\bullet}{M_{\rm disk}}\right)P,
\end{equation}
where $M_\bullet$ and $M_{\rm disk}$ are the the black hole and disk mass respectively, and P is the orbital period of a star at the inner edge of the disk \citep{rauch&tremaine1996} 
For our simulations, $t_{\rm sec} = 100P$. 

This vertical mass segregation is a form of vector resonant relaxation (VRR), which has been studied in spherical, isotropic star clusters \citep{rauch&tremaine1996, szoelgyen&kocsis2018}. In VRR orbits are reoriented by torques from other nearby orbits. However, there are notable differences from the END case. In a spherical cluster the torques add randomly, and the torque on a particular orbit scales as the square root of the number of stars in the cluster. The  timescale for VRR to significantly reorient the orbits in a cluster is

\begin{equation}
    t_{\rm VRR}\sim \sqrt{N} t_{\rm sec},
\end{equation}
where $N$ is the number of stars in the cluster. In an END, the timescale for secular vertical mass segregation is independent of $N$, as long as the mass of the disk is fixed.

We find that stars in ENDs undergo vertical mass segregation. Figure \ref{fig:nDistro_inc} shows the evolution of the out-of-plane inclination for both populations of stars in each simulation group from the $N_H$-vary set. Out-of-plane inclination $i_p$ is similar to classical inclination $i$, except it treats retrograde orbits the same as prograde orbits. The inclination of an orbit is given by

\begin{equation}
    i = \arccos{\left(\frac{j_z}{j}\right)},
\end{equation}
where $j_z$ and $j$ are the z-component and magnitude of the orbital angular momentum $\Vec{j}$, respectively. Note that $0^\circ \leq i \leq 180^\circ$. Orbits with $i < 90^\circ$ are prograde, orbits with $i = 90^\circ$ are polar, and orbits with $i > 90^\circ$ are retrograde. An orbit's out-of-plane inclination $i_p$ is related to its inclination $i$ by

\begin{equation}
    i_p = \begin{cases}
    i & i \leq 90^\circ \\
    |i - 180^\circ| & i > 90^\circ 
    \end{cases}
\end{equation}
Due to the relatively high fraction of retrograde orbits in an END (\citealt{madigan+2018}, \citealt{wernke&madigan2019}), classical inclination is not a useful measure of vertical mass segregation.  As some orbits flip to retrograde orientations, their conventional inclinations become $> 90^\circ$, and the mean inclination of the disk is inflated. Out-of-plane inclination measures the minimum angle a star's orbital plane makes with the midplane of the disk, without regard for the direction  the star is moving along the orbit.

As with radial mass segregation, vertical mass segregation shows both a strong and weak regime. In simulations with the fewest heavy stars, the heavy population shows much stronger segregation than in other simulation groups. We also explore how resistant vertical mass segregation is to changes in the initial conditions, by varying the initial inclination distribution of the heavy stars. These simulations (``$i_H$-vary'' in Table~\ref{tab:initcond}) use 25 heavy stars each, placing them in the weak mass segregation regime.  The inclination distribution is varied by changing the orbit rotation angles $\theta_a$ and $\theta_l$ (see $\S$~\ref{sec:sims}). 

Figure \ref{fig:incDistro_incBySim} shows the mean $i_p$ of each group in the ``$i_H$-vary'' set. Heavy stars drop to low inclinations very quickly, with the simulation groups converging between 100 and 150 orbital periods, which is of order the secular time. Our choice of $\Delta$ for this set ensures that the overall inclination behavior we observe will also be qualitatively valid in the strong regime, where the mass segregation will be only be more effective at dropping the inclinations of the heavy stars. Thus, in a relaxed END, we expect to find heavy stars preferentially at lower inclinations than light stars. 

\begin{figure}
    \centering
    \includegraphics[width=\columnwidth] {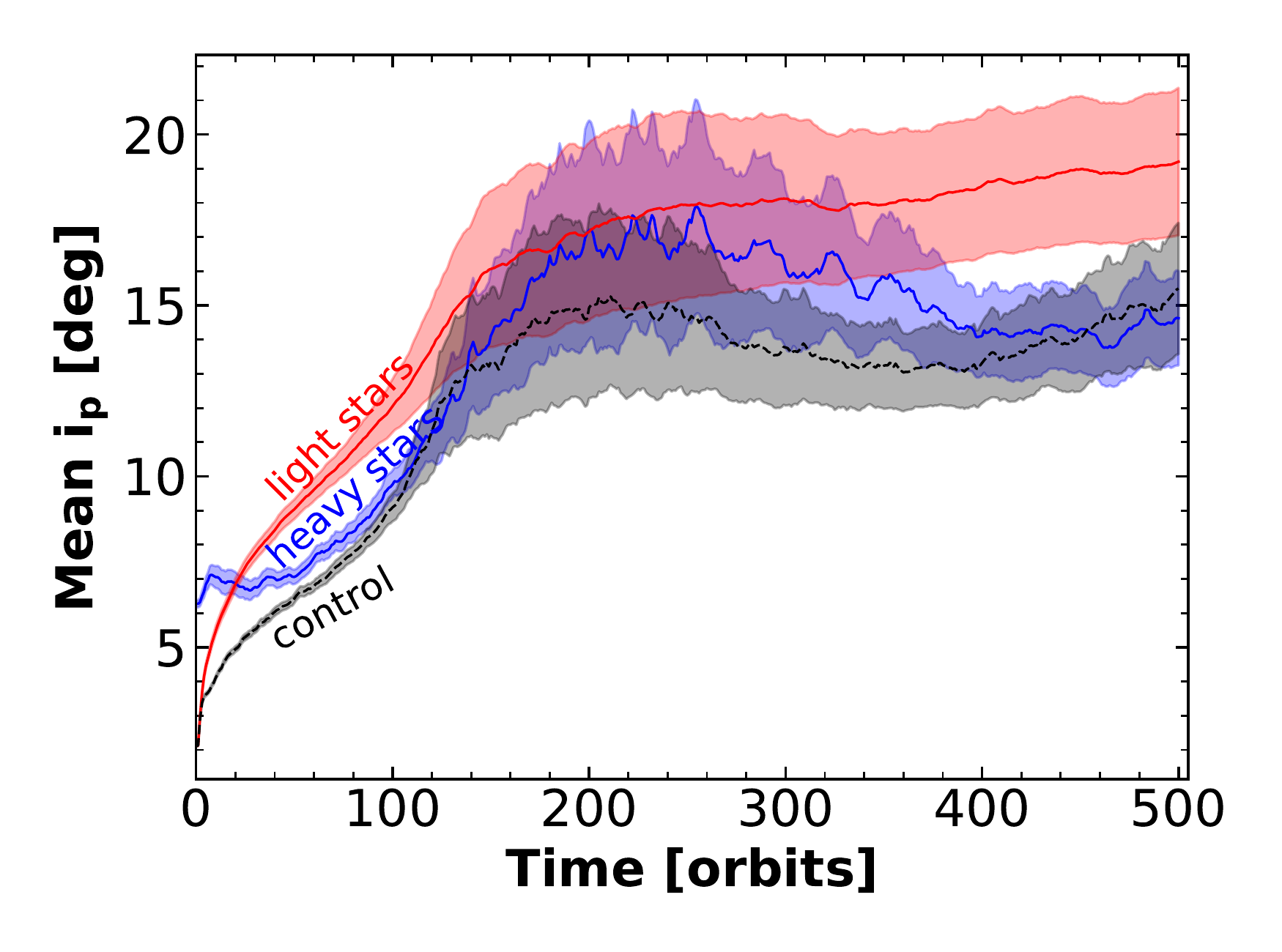}
    \caption{Mean out-of-plane inclination for the simulations with different initial heavy star inclinations (``$i_H$-vary'' from Table \ref{tab:initcond}). Each line shows the mean and standard deviation of $i_p$ for all of the stars from all of the simulations in the entire $i_H$-vary set. The blue line shows the heavy stars, the red line shows the light stars, and the dashed line shows the control simulations. The heavy stars behave more like the control stars, while they scatter the light stars to higher inclinations than they would reach without heavy stars present.}
    \label{fig:incDistro_inc}
\end{figure}

In order to more clearly show the difference between the two populations, Figure \ref{fig:incDistro_inc} condenses the different simulation groups from Figure \ref{fig:incDistro_incBySim}, showing the mean and standard deviation of $i_p$ for all stars in the entire ''$i_H$-vary'' set. This figure shows that despite starting at high inclinations, heavy stars drop to lower inclinations than the light stars on a timescale of $\sim20$ orbital periods, and remain at lower inclinations than light stars on average for the remainder of the simulation. The initial violent drop in heavy star inclinations is due to artificially high two--body relaxation (see $\S$~\ref{subsec:2body}). However, secular effects do play an important role in the inclination evolution of ENDs. 

\begin{figure}
    \centering
    \includegraphics[width=\columnwidth]{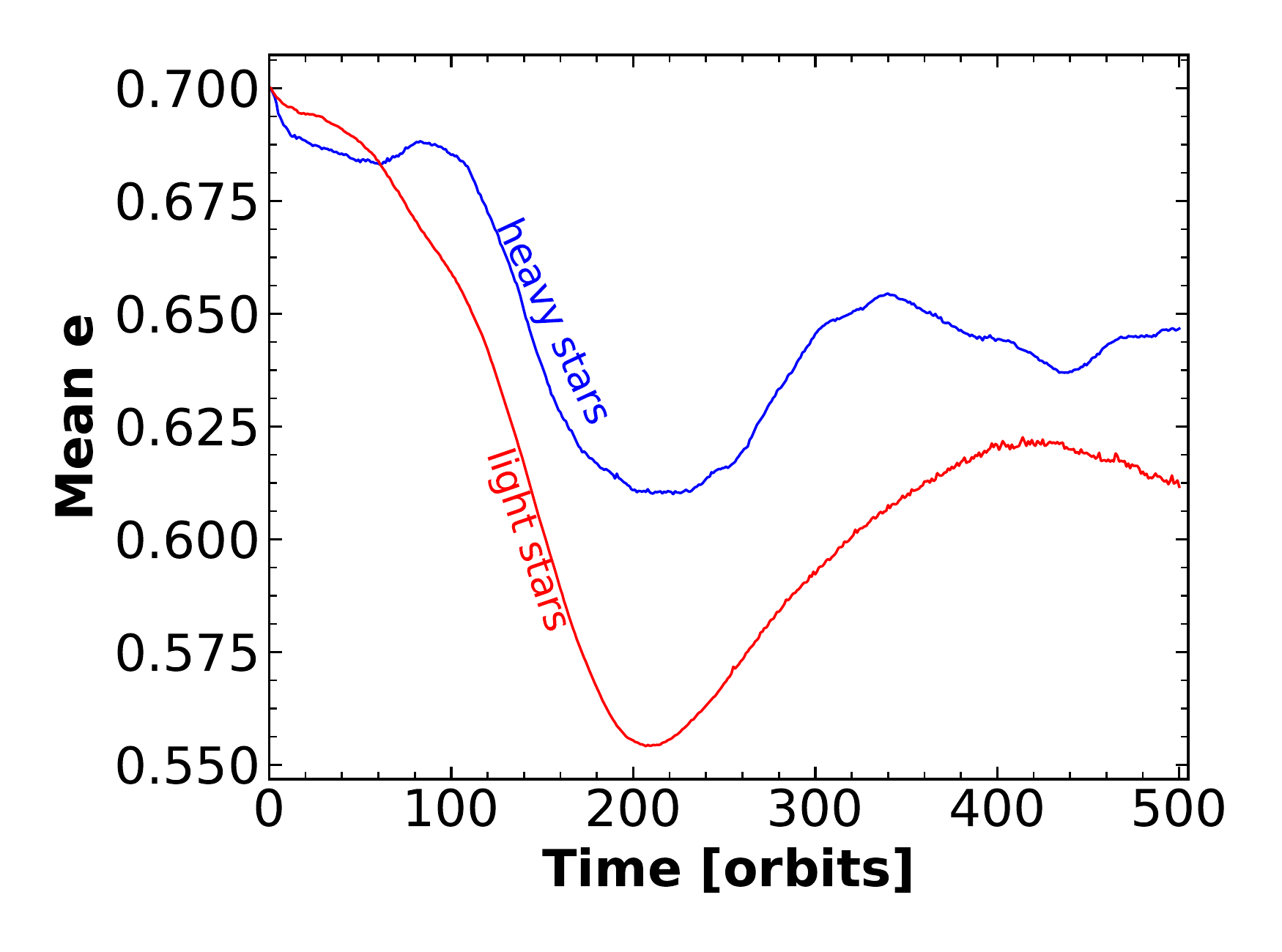}
    \caption{Mean eccentricity for the light and heavy stars in simulations that have different initial heavy star inclinations(``$i_H$-vary'' from Table \ref{tab:initcond}). Each line shows the mean of $e$ for all of the stars from all of the simulations in the entire $i_H$-vary set. As the inclination of both populations increases (Figure \ref{fig:incDistro_inc}), the eccentricity of both populations decreases after one secular time. Whereas two-body interactions would cause both inclination and eccentricity to increase, this opposite evolution of inclination and eccentricity is characteristic of secular effects.}
    \label{fig:meanE}
\end{figure}

Figure \ref{fig:meanE} shows the eccentricity evolution of both populations shown in Figure \ref{fig:incDistro_inc}. If vertical mass segregation was driven purely by two-body interactions, we would expect both the inclination and eccentricity to increase over time as the vertical 
and radial velocity dispersions increase (e.g. \citealt{ida&stewart2000}). 
Instead, as the inclination of both populations increases, the eccentricity of both populations decreases after one secular time. This opposite (and smooth) evolution of eccentricity and inclination is a result of angular momentum conservation and is characteristic of secular dynamics. This suggests that vertical mass segregation (unlike radial mass segregation, which is purely a two-body effect) is driven at least partially by orbit-averaged dynamics. 

\subsection{TDEs \label{TDErates}}

A star is tidally disrupted when it passes through its pericenter if the pericenter distance, $r_p = a(1-e)$,  is less than the tidal radius of the SMBH
\begin{equation}
    r_{\rm{t}}=\left(\frac{M_\bullet}{M_*}\right)^{1/3} R_*
\end{equation}
where $M_\bullet$ and $M_*$ are the black hole and stellar mass, respectively, and $R_*$ is the star radius \citep{rees1988}. To quantify the the TDE rate across different populations we use the disrupted fraction: the fraction of stars that are disrupted over the course of a simulation.

\begin{figure*}
    \centering
    \includegraphics{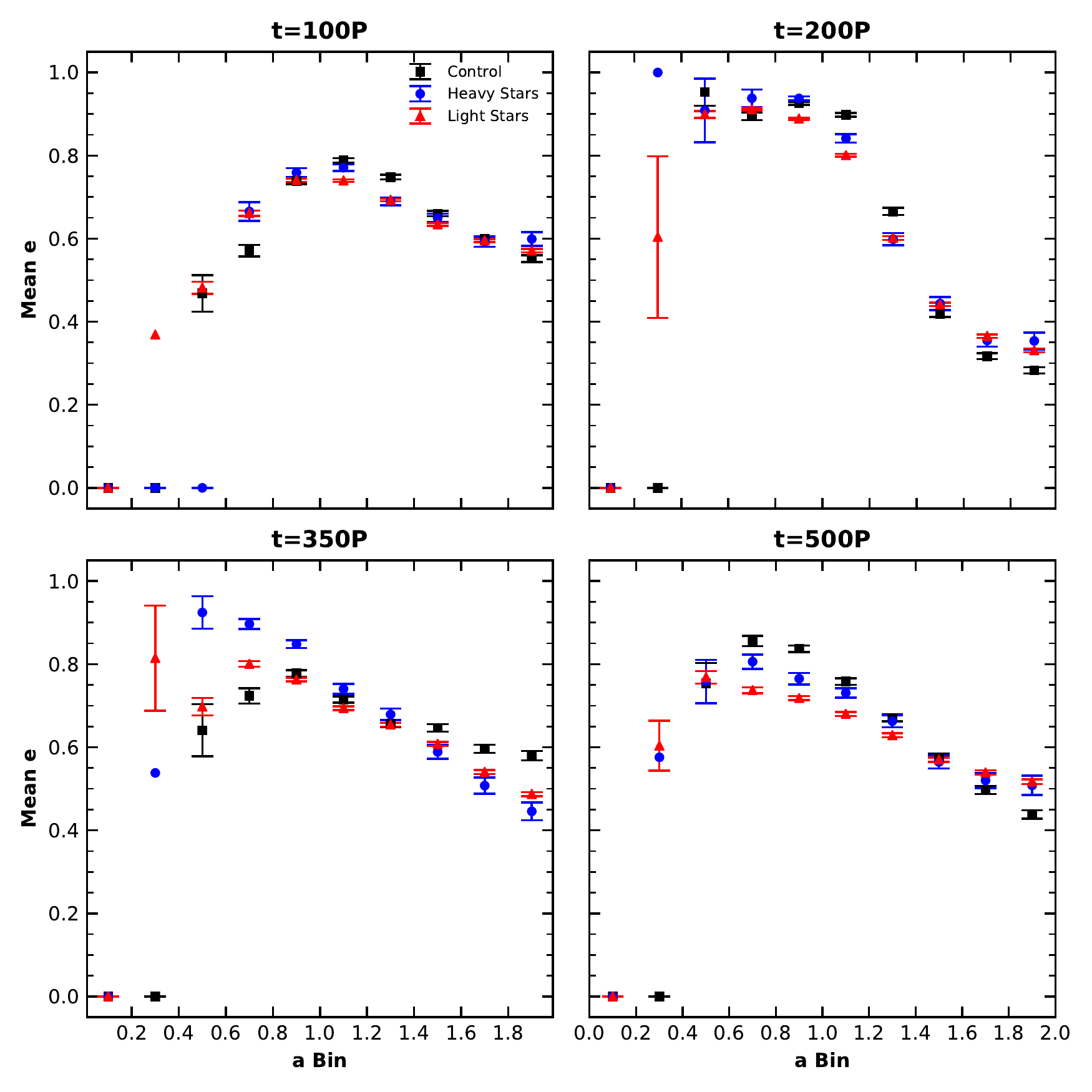}
    \caption{Development of the eccentricity gradient in simulations with different initial heavy star inclinations (``$i_H$-vary'' from Table \ref{tab:initcond}). At the beginning of the simulation, $t=0$, the eccentricity distribution of all stars is flat at $e=0.7$. Each panel shows the mean of the eccentricities of stars from all of the simulations in the $i_H$-vary set in bins of semimajor axis, broken down by population. Error bars show the standard error of the mean. Bins that show an eccentricity of 0 contain no stars at the selected time. \textit{Top Left Panel}: By one secular time, the eccentricity gradient has started to take shape. \textit{Other Panels}: Over the course of the simulation, the gradient changes shape slightly, but the heavy stars always have higher eccentricities than the light stars on average in bins where the bulk of TDEs come from, between a=0.6 and a=1.2.}
    \label{fig:eGrad}
\end{figure*}

Stars at the inner edge of the disk are preferentially disrupted, as they have lower angular momentum due to the negative eccentricity gradient in ENDs and lower angular momentum stars are more easily torqued to a high eccentricity \citep{madigan+2018}. Therefore, we should expect to see a larger fraction of heavy stars disrupting than light stars, as they are more central concentrated due to radial mass segregation.

Vertical mass segregation is important too, dropping the heavy stars to lower inclinations than the light stars. For low inclination orbits, the torque from the disk is nearly aligned with their angular momentum vectors, and changes their eccentricity rather than their orbital orientation. Thus, heavy stars are raised to higher eccentricities than the light stars even at a fixed semimajor axis. This can be seen in Figure~\ref{fig:eGrad}. This figure shows the development of the eccentricity gradient in eccentric disks with and without a mass spectrum. While the shape of the gradient changes over the course of the simulations, the heavy stars always have higher mean eccentricities than light stars between semimajor axes of 0.6 and 1.2. The vast majority of orbits that lead to tidal disruption also have semimajor axes in this range, confirming that heavy stars should be easier to disrupt than light stars at any given semimajor axis.

\begin{figure}
    \centering
    \includegraphics[width=\columnwidth]{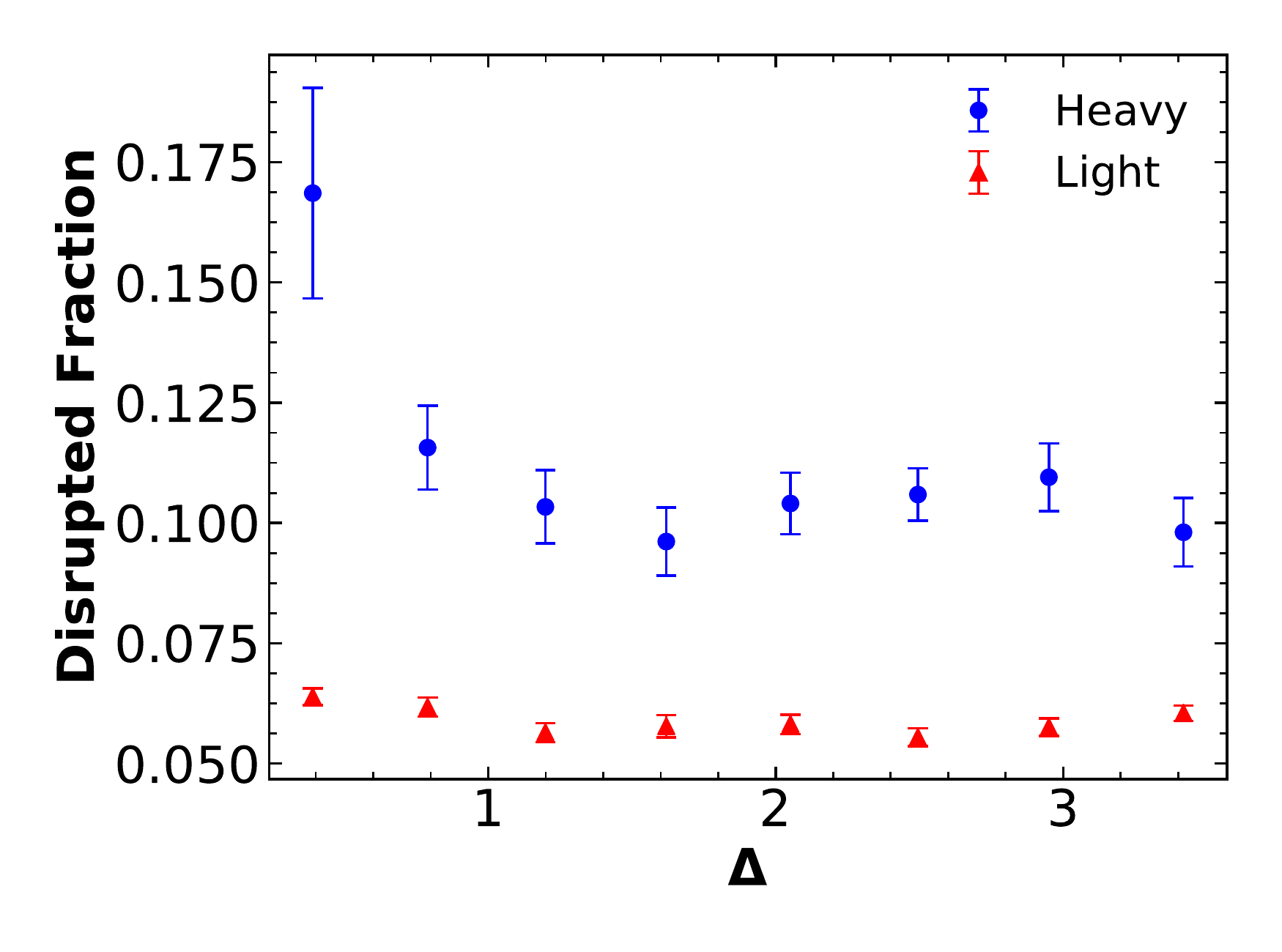}
    \caption{Disrupted fraction vs $\Delta$ (or mass segregation strength, lower $\Delta$ is stronger mass segregation) for simulations with different numbers of heavy stars (``$N_H$-vary'' from Table \ref{tab:initcond}). Each point shows the mean disrupted fraction for all simulations in a group, broken down into heavy and light stars. Error bars show the standard error of the mean. The disruption rate for light stars is independent of $\Delta$, while heavy star disruption rate increases at low $\Delta$ (due to stronger mass segregation).}
    \label{fig:nDistro_TDE}
\end{figure}

Figure \ref{fig:nDistro_TDE} shows the disrupted fraction for both populations in simulation groups with different numbers of heavy and light stars (``$N_H$-vary'' in Table~\ref{tab:initcond}). Fewer heavy stars (lower $\Delta$) translates to stronger mass segregation, and the group's heavy stars being closer to the inner edge of the disk at higher eccentricity. In all cases, disrupted fraction of the heavy stars is larger than that of the light stars as expected. In particular, the heavy stars in the most strongly segregated group with $\Delta=0.384$ have the smallest mean semimajor axis and are the most likely to be disrupted. 

\begin{figure}
    \includegraphics[width=\columnwidth]{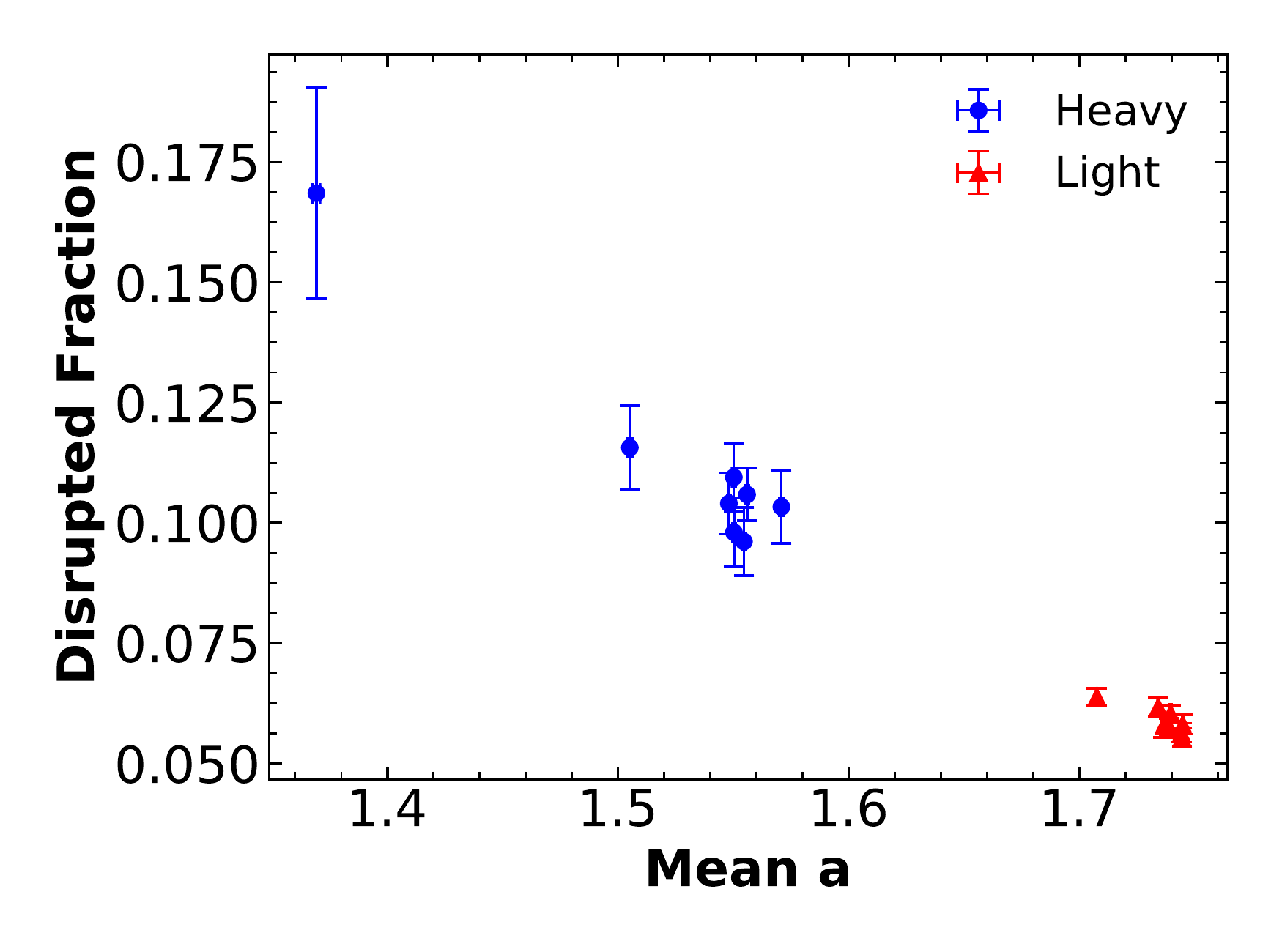}
    \caption{Disrupted fraction vs mean semimajor axis for simulations with different numbers of heavy stars (``$N_H$-vary'' from Table \ref{tab:initcond}). Horizontal axis values are obtained by taking the mean of all stars in all simulations from a group during all timesteps after the first 100 orbital periods. We consider times after 100 orbital periods to allow the disk a full secular time to relax. Horizontal error bars show the standard error of this mean, and are generally smaller than the points. Populations spending more time at low semimajor axes have a correspondingly higher disrupted fraction.
    }

    \label{fig:nDistro_TDEvsma}
\end{figure}

Figure \ref{fig:nDistro_TDEvsma} again shows the disrupted fraction for each population from the $N_H$-vary set (similarly to Figure \ref{fig:nDistro_TDE}), but now as a function of the time- and star-averaged semimajor axis of the population.\footnote{Time averages are taken between $t=100P$ and the end of the simulation. This is done to allow the disk a secular time to relax, as the TDEs do not begin to happen until after one secular time.} Populations with a lower mean semimajor axis have a correspondingly higher disrupted fraction.

\begin{figure}
    \includegraphics[width=\columnwidth]{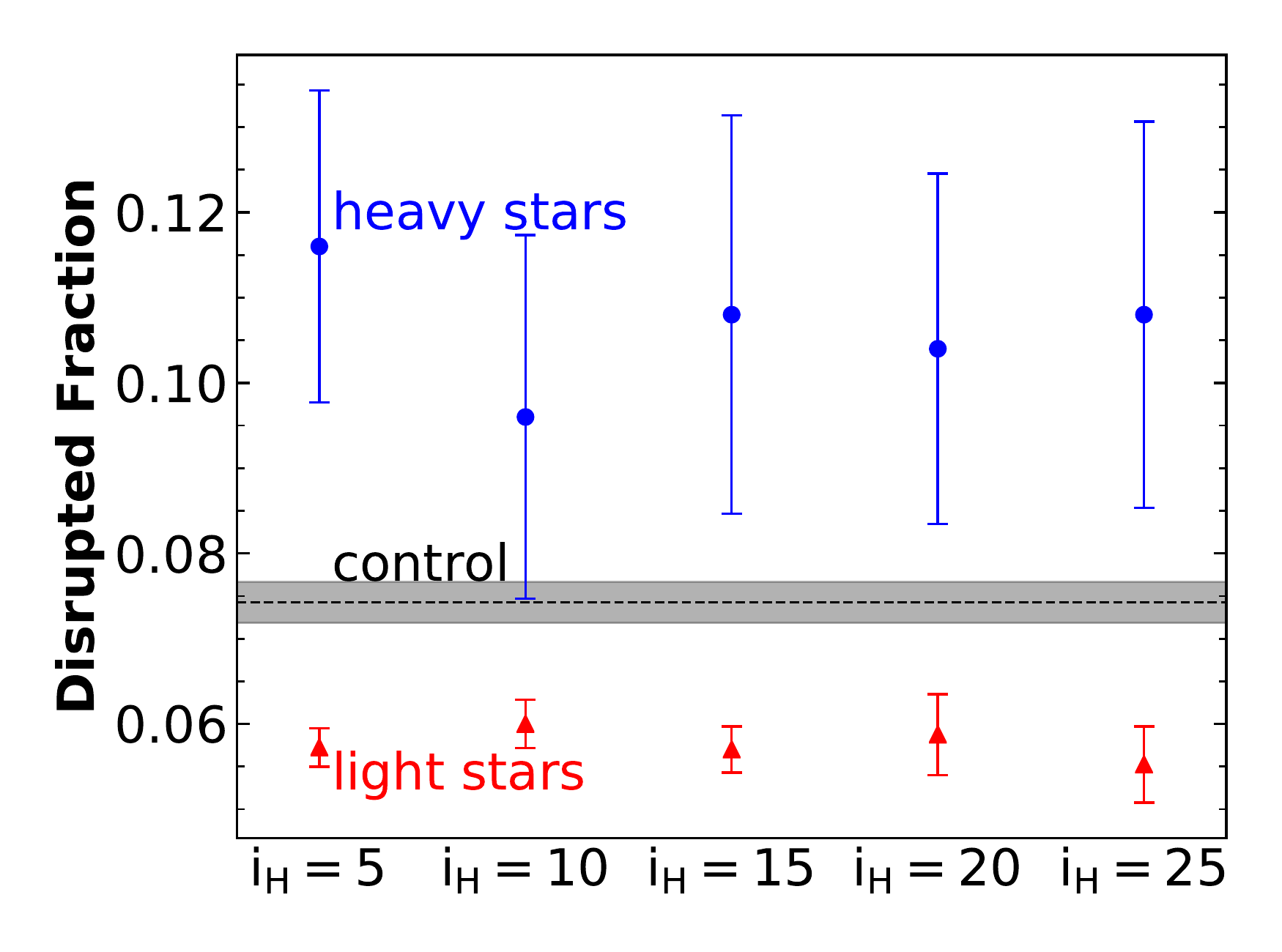}
    \caption{Disrupted fraction for simulations with different initial heavy star inclinations (``$i_H$-vary'' in Table~\ref{tab:initcond}). Each point shows the mean of the disrupted fraction from the simulations in the group, with heavy stars shown in blue and light stars shown in red. The errorbars are the standard error of the mean for each simulation group. The dashed line shows the TDE rate per star of the control simulations. The TDE rate is a weak function of the initial inclination, with the heavy star disrupted fraction always exceeding the light star disrupted fraction.
    }
    \label{fig:incDistro_TDE}
\end{figure}

Figure~\ref{fig:incDistro_TDE} shows the disrupted fraction of each population for simulations with different initial heavy star inclinations (``$i_H$-vary'' in Table~\ref{tab:initcond}). Once again the fraction of heavy stars that are disrupted is larger than the fraction of light stars. There is no significant difference between the groups because they all have the same $\Delta$, and show the same degree of vertical mass segregation. In particular, the heavy stars from different simulation groups have all reached the same inclination by the time the TDEs begin to occur. 

To summarize, in order to be disrupted, a star's orbit must have a torque applied to it to raise it to very high eccentricity. Stars at lower semimajor axes should be easier to disrupt because they need less torque due to their higher equilibrium eccentricity, and stars at low inclination should be easier to disrupt because the torque from the disk is more aligned with their angular momentum vectors, and so changes their orbital eccentricities rather than their orientations. 
Radial mass segregation places heavy stars at low semimajor axes, and vertical mass segregation places stars at low inclination. Across all of our simulations, a larger fraction of heavy stars are disrupted, as we would expect. 

 Mass segregation can also increase the specific TDE or capture rates of heavy objects in isotropic clusters.
 For example, \citet{vasiliev2017} finds that the disruption rate of $1 M_{\odot}$ stars is $\sim 120$ times larger than the capture rate of $10 M_{\odot}$ black holes in a two component, isotropic Fokker--Planck model for the Galactic Center.\footnote{The disruption and capture rates are $\sim6\times 10^{-5}$ and $5\times 10^{-7}$ per year.  \citet{vasiliev2017} gives $5\times 10^{-6}$ for the latter, but this is actually the mass of black holes (in solar units) that are consumed per year (E. Vasiliev, personal communication).} Considering that there are $\sim 300$ times more of the former within the gravitational influence radius of this model, the specific capture rate of black holes is a factor of $\sim$2.5 greater than the specific TDE rate of solar mass stars.  Coincidentally, this is similar to the enhancment we find in ENDs.

\subsection{Two--body relaxation}
\label{subsec:2body}
Due to computational constraints our simulations contain fewer and more massive stars than a physical END, which would artificially decrease the two--body relaxation time. This means mass segregation would occur faster in our simulations than in a realistic system. In particular, while the mass segregation time scale is shorter than the secular time scale in our simulations, the ordering of these two time scales could be flipped in reality. As previously discussed ENDS start to produce TDEs after one secular time. If the disk has not mass segregated within this time, the heavy star TDE rate may not be enhanced relative to the light star TDE rate in young ENDs, as we found in $\S$~\ref{TDErates}. 

We use the one dimensional model of \citet{alexander+2007} to quantify the mass segregation time scale in disks with realistic numbers of stars. For completeness the relevant equations are summarized in appendix~\ref{app:alexander}. This model contains two populations of different masses, and the velocity dispersion of each species evolves due to two--body encounters. We fix the disk mass to be one percent of the central mass, as in our N--body simulations.

\begin{figure}
    \centering
    \includegraphics[width=\columnwidth]{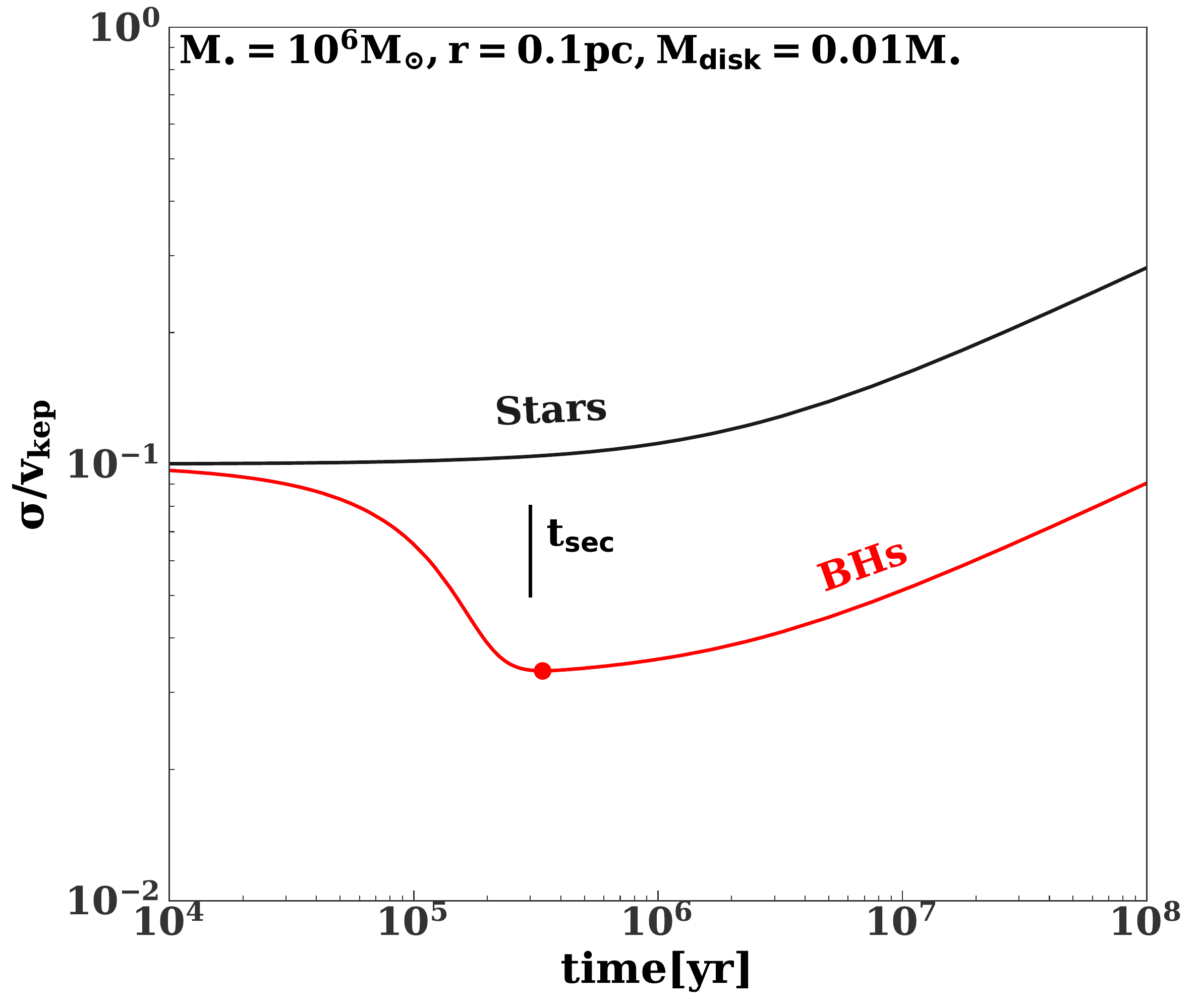}
    \includegraphics[width=\columnwidth]{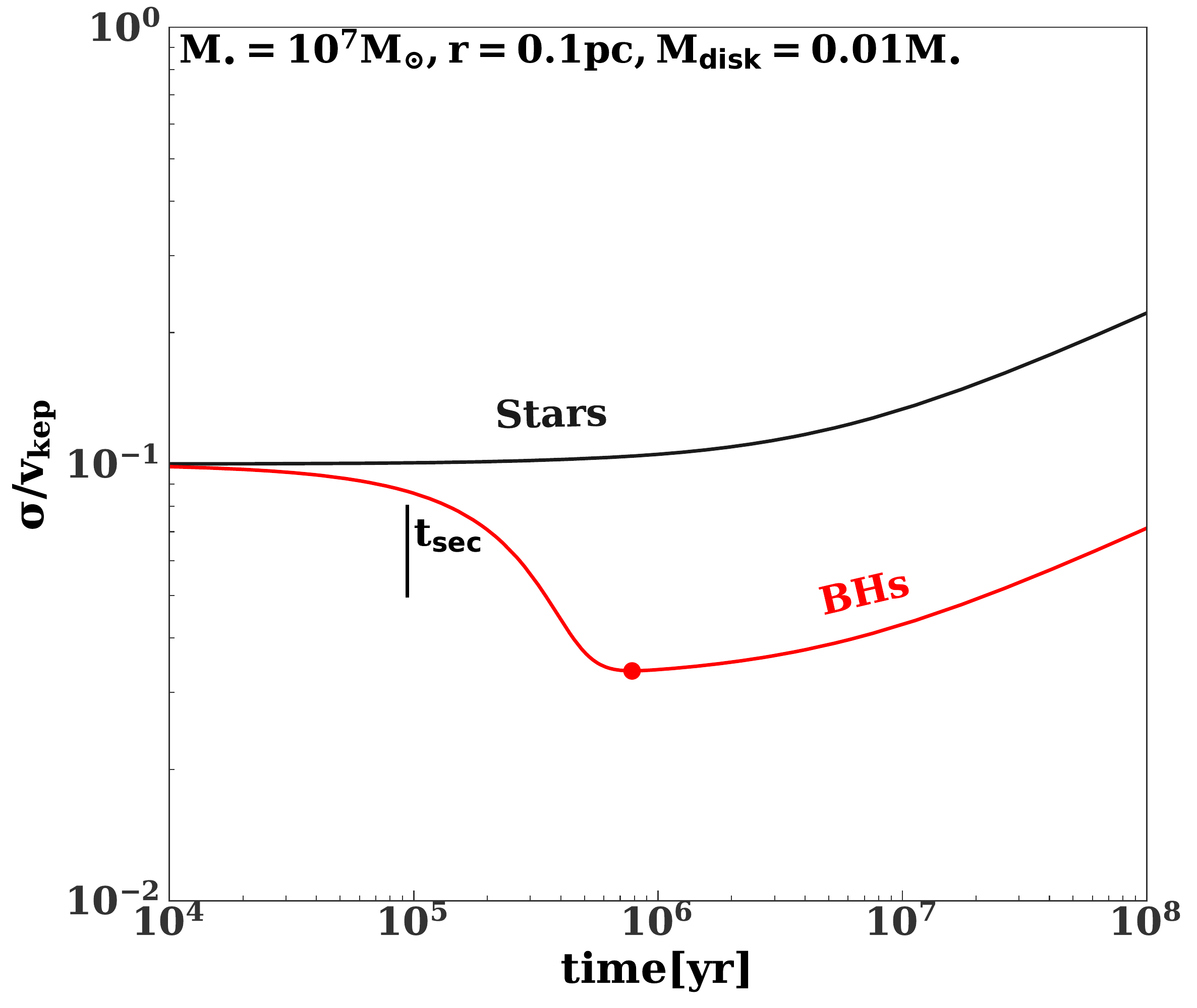}
    \caption{Velocity dispersion as a function of time for disks around different mass SMBHs from the one--dimensional model of \citet{alexander+2007} (see text for details). 
    The disk contains $1 M_{\odot}$ (\emph{black lines}) and $10 M_{\odot}$ stars (\emph{red lines}), with 1000 times more of the former. In both panels all stars start with a velocity dispersion equal to ten percent of the local Keplerian velocity.  The black, vertical line in each panel indicates the secular time, while the red dots indicate the mass segregation time.
    For larger SMBHs, the mass segregation timescale in an END is longer than the secular time, and TDEs would occur before the disk has mass--segregated. There would be no relative enhancement in the heavy star TDE rate at early times in this case.
    \label{fig:alexanderExample}}
\end{figure}

\begin{figure}
    \centering
    \includegraphics[width=\columnwidth]{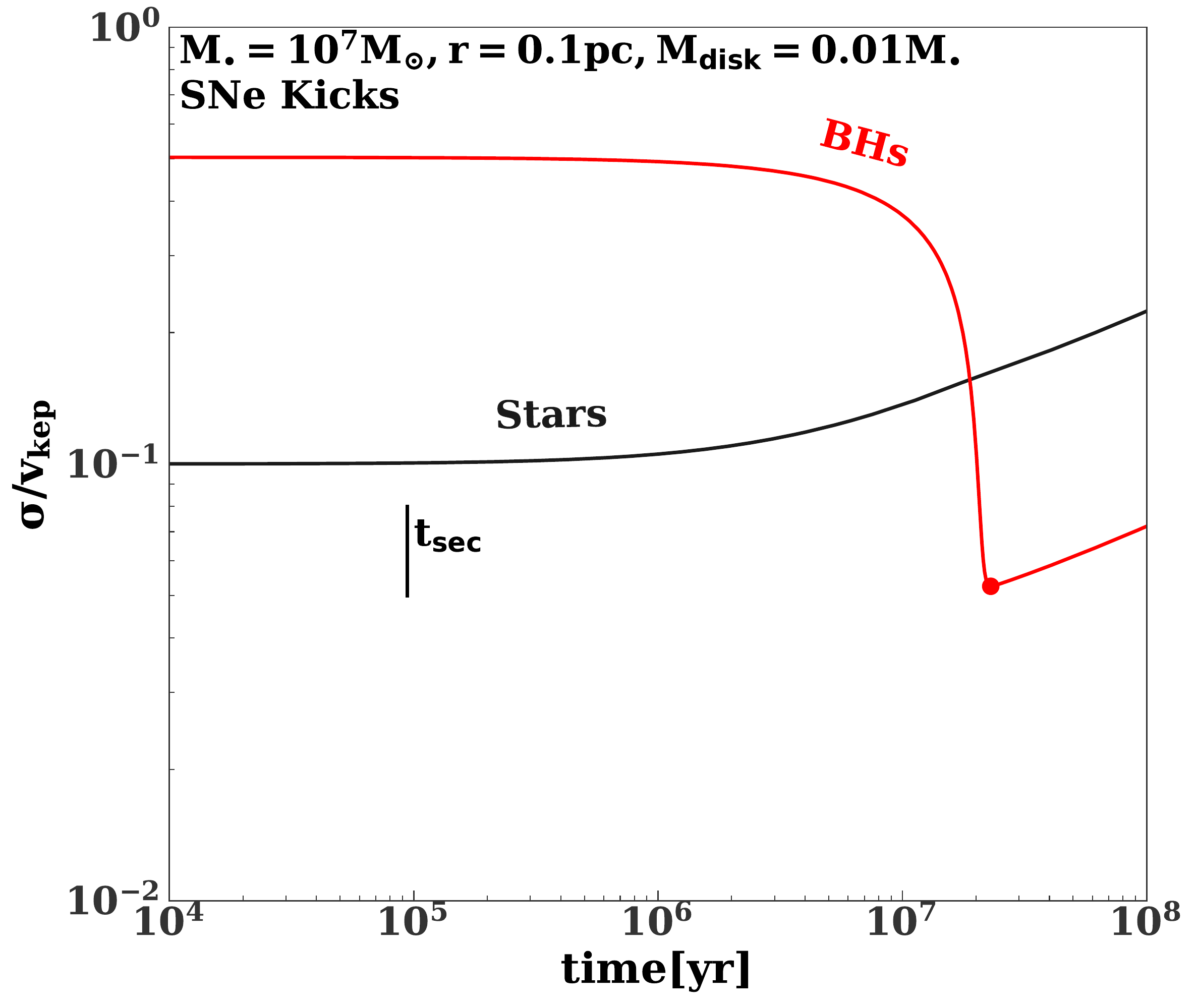}
    \caption{Same as the bottom panel Figure~\ref{fig:alexanderExample}, except we add $265$ km s$^{-1}$ to the velocity dispersion of the heavy stars to simulate the effects of supernova kicks. Supernova kicks increase the velocity dispersion of compact objects, which would reduce the rate of close encounters between compact objects and the central SMBH.
    \label{fig:alexanderExample2}}
\end{figure}

Figure~\ref{fig:alexanderExample} shows a couple of illustrative examples of the evolution of the velocity dispersions in this model. The top panel shows a case where the mass segregation time is comparable to the secular time. (We define a ``mass segregation time'' as the point where the velocity dispersion of the heavy stars is minimized; this is comparable to the timescale for the two species to reach energy equipartition).
In the bottom panel of Figure~\ref{fig:alexanderExample} the SMBH mass is increased, so that the heavy stars segregate after one secular time.
Finally, in Figure~\ref{fig:alexanderExample2} the heavy stars start with a higher velocity dispersion than the light stars to simulate the effect of supernova kicks, making the mass segregation time much longer.

We find the mass segregation time in this model is well approximated by
\begin{equation}
t_{\rm seg}\approx 3.2\times 10^{5} {\rm yr} \left(\frac{M_\bullet}{10^6 M_{\odot}}\right)^{0.38}  \left(\frac{\sigma_o}{0.1 v_{\rm kep}}\right)^{3.7}
 \left(\frac{r}{0.1 {\rm pc}}\right)^{1.5},
\label{eq:massSeg}
\end{equation}
where $\sigma_o$ is the initial velocity dispersion of both species and $v_{\rm kep}$ is the local Keplerian velocity. The mass segregation time is a weak function of the relative number of heavy and light stars in the disk. On the other hand the secular time is 

\begin{equation}
t_{\rm sec}\approx 2.9\times 10^{5} {\rm yr} \left(\frac{M_\bullet}{10^6 M_{\odot}}\right)^{-0.5}
 \left(\frac{r}{0.1 {\rm pc}}\right)^{1.5}.
 \label{eq:tsec}
\end{equation}
The mass segregation time will be shorter than the secular time as long as 

\begin{equation}
    M_{\bullet}<9\times 10^5 M_{\odot} \left(\frac{\sigma_o}{0.1 v_{\rm kep}}\right)^{-4.9}.
    \label{eq:tms-tec}
\end{equation}
To summarize, we expect dynamically cold ENDs around low mass SMBHs to mass segregate before one secular
time, when they would begin to produce TDEs. In this regime, we expect a relative enhancement in the heavy 
star TDE rate at all times as described in $\S$~\ref{TDErates}. For more massive SMBHs, the mass segregation time is longer,
and there would would be no relative enhancement in the heavy star TDE rate at early times.Although the black hole mass function remains poorly constrained below $10^6 M_{\odot}$, there are 
several SMBHs with credible mass measurements of a few$\times 10^5 M_{\odot}$ \citep{greene+2019}. Also,
scaling relations suggest that some TDE hosts have SMBH masses below $10^6 M_{\odot}$ \citep{wevers+2019}.

\section{Discussion \label{sec:discussion}}

In this section, we present a discussion of our results in a wider astrophysical context. We discuss how realistic variations of the tidal radius between different stellar species would affect our results. Additionally, we speculate on the mass function of a realistic END, including likely $\Delta$ values and the resulting mass segregation and structure of the disk. Lastly, we consider the effect that a different initial eccentricity would have on TDE rates.

\subsection{Effects of stellar properties on encounter rates \label{subsec:rt}}

\begin{figure}
    \centering
    \includegraphics[width=\columnwidth]{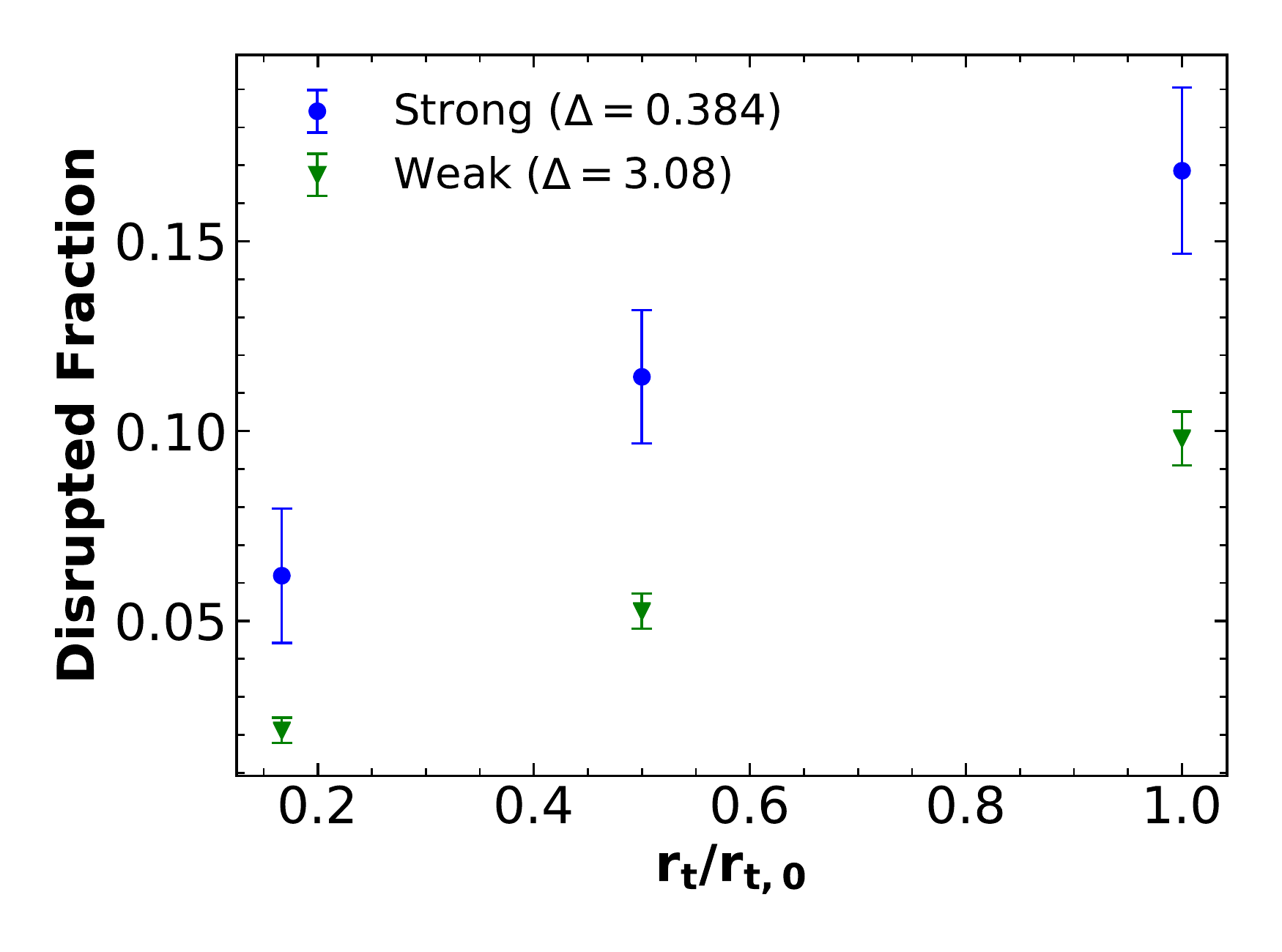}
    \caption{Mean disrupted fraction of heavy stars vs. tidal radius, for two simulation groups, one in each mass segregation regime from the $N_H$-vary set. $r_{t, 0}$ is the tidal radius of a light star. The disrupted fraction is linear with the tidal radius in both cases.}
    \label{fig:rt}
\end{figure}

Both heavy and light stars have the same tidal radius in our simulations. In reality, the tidal radius depends on quantities such as the stellar mass, radius, and spin \citep{rees1988, golightly+2019}. Also, compact objects (e.g. stellar mass black holes) will not be tidally disrupted by the central SMBH, although they can be captured if they pass too close to it. Specifically, objects with a Keplerian pericenter, $r_{\rm capt}$, satisfying\textbf{}

\begin{equation}
    r_{\rm capt}\leq \frac{8 G M_\bullet}{c^2}
\end{equation}
would plunge into the horizon of a Schwarzschild black hole \citep{gair+2006}.\footnote{The corresponding relativistic pericenter is $4 \frac{G M}{c^2}$ in Boyer--Lindquist coordinates.} ENDs are generally in the full loss cone regime, meaning close encounters between disk objects and the central SMBH have a uniform pericenter distribution \citep{wernke&madigan2019}. Therefore, the cross section for disruption (capture) is expected to scale linearly with the disruption (capture) radius. By screening out heavy TDEs with pericenters above other values of $r_t$, we can see how the disrupted fraction of heavy objects changes with the tidal radius. Figure~\ref{fig:rt} shows that the disrupted fraction indeed scales linearly with the tidal radius in our simulations.

Thus, 
\begin{equation}
    \frac{f_{\rm capt}}{f_{\rm dis}}=\frac{\chi \frac{8 G M}{c^2} }{\left(\frac{M_\bullet}{M_\star}\right)^{1/3} R_\star(M_\star)},
\end{equation}
where $f_{\rm dis}$ ($f_{\rm capt}$) is the fraction of stars that are disrupted (fraction of stellar black holes that are captured); 
$M_\star$ and $R_\star$ are the stellar mass and radius; $\chi$ is the relative enhancement in the black hole capture rate due to
mass segregation. $\chi\approx 2-3$ in our simulations (see Figure~\ref{fig:incDistro_TDE}), and increases as the relaxational 
coupling parameter ($\Delta$) decreases.

If all of the stars are Sun--like, this ratio is 
\begin{equation}
    \frac{f_{\rm capt}}{f_{\rm dis}}\approx \frac{1}{3} \left(\frac{M_\bullet}{10^6 M_\odot}\right)^{2/3}.
\end{equation}
This exceeds unity for $M_\bullet\gsim 5.2 \times 10^6 M_{\odot}$. For lower mass SMBHs the fraction of compact objects that are captured is smaller than the fraction of stars that are disrupted, as the effects of mass segregation cannot compensate for the smaller capture radius.
Among main sequence stars in an END, the specific TDE rate\footnote{The disruption rate per star.} would be an increasing function of stellar mass due to mass segregation and the larger radii of high mass stars. For $M_\star \gsim 1 M_{\odot}$, the stellar radius scales approximately as the square root of the stellar mass, implying that
\begin{equation}
r_t=\left(\frac{M_\bullet}{M_\star}\right)^{1/3} R_\star(M_\star)\propto M_\star^{0.2},
\end{equation}
along the upper main sequence. Despite their higher specific TDE rate, the disrupted fraction may decrease with stellar mass, as the stellar lifetime increases. However, this depends on the timescale over which an END could sustain highly eccentric orbits. Disruptions and captures remove objects from the disk and lower its mean eccentricity, so that the TDE rate decreases with time and may eventually drop to zero \citep{madigan+2018}. Unfortunately, this timescale is poorly constrained by existing simulations, which have artificially high two--body relaxation times. If it is shorter than the lifetime of a massive star, the disrupted fraction would increase with stellar mass. In this case, close encounters between disk black holes and the central SMBH would be virtually nonexistent (as the disk would not be able to excite disk black holes to highly eccentric orbits by the time they form).

\subsection{END Mass Functions \label{subsec:pmf}}

A real END would of course contain far more than two stellar species, but placing constraints on a likely present-day mass function (PMF) is very difficult. As discussed by AH09, the two-species approximation we use here falls naturally out of ``universal'' initial mass functions (IMFs) such as those developed by \citet{salpeter1955}, \citet{miller&scalo1979}, and \citet{kroupa2001}. Evolved populations from these IMFs generally have $\Delta < 0.1$ (AH09), placing them firmly in the strong segregation regime. \citet{merritt2013} shows that a population evolving from the Kroupa IMF reach $\Delta \approx 0.05$, again in the strong regime. 

However, there is also evidence to suggest that star formation and IMFs in galactic nuclei near SMBHs may be different from the universal IMFs used for field stars (e.g. \citealt{levin&beloborodov03}; \citealt{milosavljevic&loeb2004}; \citealt{paumard+2006}; \citealt{levin2007}; \citealt{bartko+2010}; \citealt{lu+2013}).

ENDs in particular should also have different IMFs depending on whether the disk stars were formed on eccentric orbits, or if the disk was formed dynamically (e.g. during a galactic merger) after the stars were formed. ENDs formed dynamically from stellar populations with universal IMFs would have $\Delta < 0.1$ as previously discussed, placing them in the strong segregation regime. Conversely, ENDs formed from an initially eccentric thin gas cloud around the SMBH could have top--heavy mass functions. \citet{alexander+2008} used hydrodynamical simulations to study the behavior of initially eccentric accretion disks. They found that smaller, loosely-bound clumps that formed in the disk were particularly vulnerable to tidal disruption at pericenter, leading to only the high-mass, dense clumps surviving. Extending their results to star formation, a top--heavy mass function of clumps in a gas disk would likely lead to a top--heavy stellar IMF. A detailed exploration of exactly what $\Delta$ would fall out of such a top--heavy mass function is beyond the scope of this work, however we can predict that it would be higher than that of a universal IMF, such that $\Delta > 0.1$. Thus, ENDs formed from an initially eccentric gas disk would likely be more weakly segregated than ENDs formed dynamically from existing stars. 

The present-day mass function (PMF) of an END would depend on a myriad of factors, including the formation history of the disk, which would influence the IMF. The age of the disk is also very important, as it affects the PMF through stellar evolution and loss of disk stars to TDEs and captures over time.

Lastly, we note that for most reasonable mass functions, there are many more light stars than heavy stars \citep{kroupa+2013}. Thus, while our results suggest a factor of 2-3 enhancement in the \textit{specific} TDE (capture) rate of heavy stars (compact objects), TDEs from light stars will still dominate the overall TDE rate of the END.

\subsection{The Effect of Eccentricity \label{subsec:eccentricity}}

\begin{figure}
    \centering
    \includegraphics[width=\columnwidth]{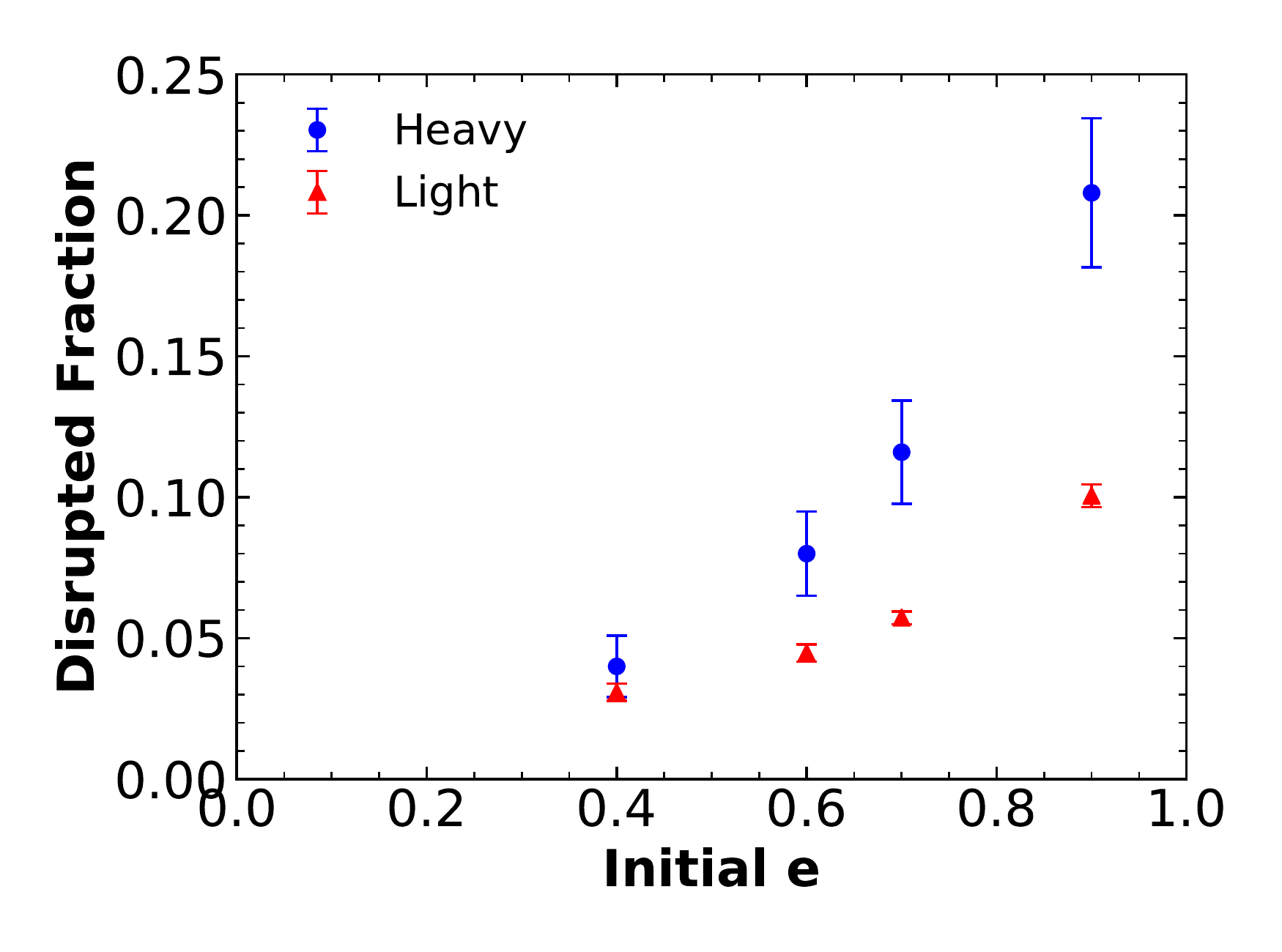}
    \includegraphics[width=\columnwidth]{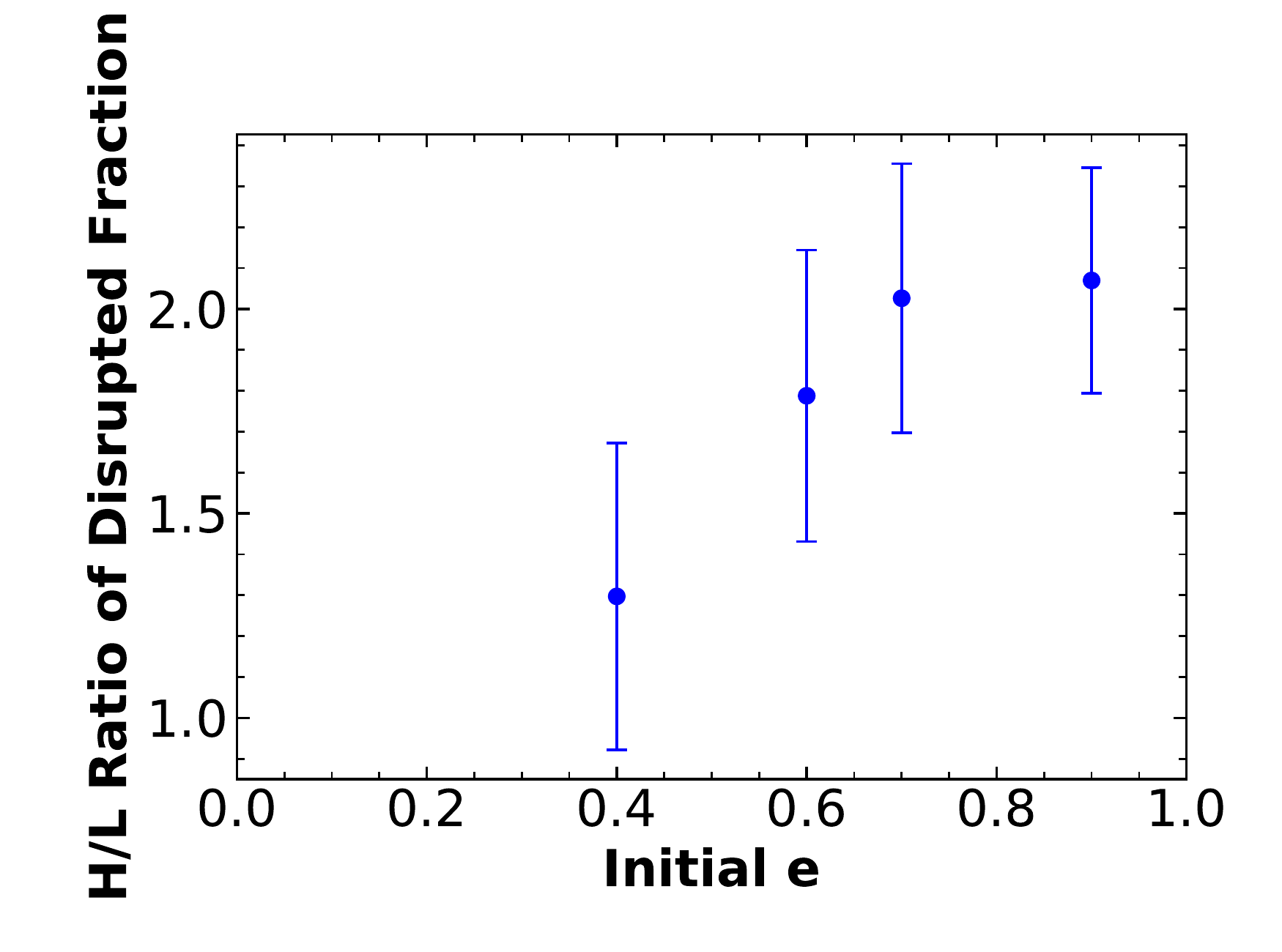}
    \caption{Disrupted fraction for simulations with different initial eccentricities. \textit{Top Panel}: Each point shows the mean of the disrupted fraction from the simulations in the group, with heavy stars shown in blue and light stars shown in red. The errorbars are the standard error of the mean for each simulation group. The disrupted fraction of both species is correlated with the initial eccentricity of the disk orbits. \textit{Bottom Panel}: The ratio of the heavy disrupted fraction to light disrupted fraction vs. the initial eccentricity. The enhancement of the heavy specific TDE rate is also correlated with the initial eccentricity.}
    \label{fig:eTDEs}
\end{figure}

So far, we have only considered a delta function for the initial eccentricity distribution with $e=0.7$. Here, we briefly discuss the effect that changing the initial eccentricity has on TDE rates.

In principle, increasing the mean eccentricity of the disk should increase the TDE rate for both species, as the disk as a whole has less angular momentum. Similarly, decreasing the mean eccentricity of the disk should decrease the TDE rates for both species.

In order to test this prediction, we use a third small set of simulations. In this third set, we start both species with the same $\Delta$ and inclination distribution, but change the initial eccentricity of the disk orbits. Each simulation contains 25 heavy stars for easy comparison with the $i_H$-vary set, and there are ten individual simulations for each eccentricity considered.

The top panel of Figure \ref{fig:eTDEs} shows the mean disrupted fraction for both species as a function of initial eccentricity for this set. The disrupted fraction for both species is an increasing function of the initial eccentricity as expected. This trend is consistent with \citet{madigan+2018}, who found that as an END loses mass through disrupted stars, the average eccentricity of the disk drops and the TDE rate is suppressed. The bottom panel of Figure \ref{fig:eTDEs} shows that the enhancement of the heavy star disrupted fraction increases weakly with the initial eccentricity.
\section{Summary \label{sec:conclusion}}

In this paper, we presented the first study of an eccentric nuclear disk with two stellar species. Here, we give a summary of our results and their implications. 

\begin{enumerate}

    \item 
    \textit{Radial Mass Segregation}: Similar to previous studies with isotropic clusters and axisymmetric disks, two-species ENDs undergo mass segregation in energy/semimajor axis space. The strength of the mass segregation falls into two regimes, determined by the relaxational coupling parameter $\Delta$. The cutoff between the strong and weak segregation occurs around $\Delta \sim 1$.
    
    \item 
    \textit{Vertical Mass Segregation}: In a two-species eccentric disk, heavy stars sink to lower inclinations than light stars on average. This process is highly resistant to artificially raising the inclinations of the heavy stars (e.g. with supernova kicks) in our simulations with artificially strong two-body relaxation. In a real disk, this process may be significantly delayed by supernova kicks. As with radial mass segregation, vertical mass segregation has both a weak and strong regime. 
    
    \item 
    \textit{TDE rates}: The negative eccentricity gradient in stable eccentric disks causes stars at low semimajor axes to have higher equilibrium eccentricities, where they are more easily driven onto orbits that take the star within the tidal radius of the SMBH. Stars at low inclinations have their eccentricities altered by secular torques rather than their orientations, leading to these orbits reaching higher eccentricities where they are more susceptible to disruption. Heavy stars are preferentially found at low semimajor axes and low inclinations due to mass segregation, and are more likely to be disrupted than light stars.

    We find that mass segregation can increase the specific TDE rate of heavy stars in an END by a factor of 2--3 relative to light stars, assuming the same tidal radius for both species. Due to the much larger number of light stars, TDEs from light stars will still dominate the overall TDE rate of the END. In a real system, the ratio of the heavy to light star specific TDE (or capture) rate depends on the age of the disk and the mass of the central SMBH. In particular, this enhancement will be present for young, dynamically cold ENDs around $\lesssim 10^6 M_{\odot}$ SMBHs (see eq.~\ref{eq:tms-tec}). The larger tidal radii of massive stars will further enhance their TDE rate, considering that ENDs are typically in the full loss cone regime \citep{wernke&madigan2019}. This enhancement from larger stellar and tidal radii will be much weaker in isotropic clusters, which are at least partially in the empty loss cone regime \citep{macleod+2012, kochanek2016}. However, mass segregation could still result in a factor of a few increase in the specific capture rates of heavy objects in an isotropic cluster \citep{vasiliev2017}.
    
\end{enumerate}

Finally, we note that mass segregation may explain the observed wavelength dependence of the orientation of the END in M31 \citep{lockhart+2018}. However, a broader range of stellar masses and stellar evolution would have to be included in our model before we could make detailed comparisons with these observations. 

\acknowledgments
We thank Nicholas Stone, Eugene Vasiliev, and the anonymous referee for helpful comments.
We gratefully acknowledge support from NASA Astrophysics Theory Program under grant NNX17AK44G. 
Simulations in this paper made use of the \texttt{REBOUND} code which can be downloaded freely at \url{https://github.com/hannorein/rebound}. 
This work utilized the RMACC Summit supercomputer, which is supported by the National Science Foundation (awards ACI-1532235 and ACI-1532236), the University of Colorado Boulder, and Colorado State University. The Summit supercomputer is a joint effort of the University of Colorado Boulder and Colorado State University.

\software{\texttt{REBOUND} \citep{rein&liu2012}} 

\clearpage
\appendix 
\section{One dimensional model for dynamical evolution of disks}
\label{app:alexander}

In $\S$~\ref{subsec:2body}, we solve for the velocity dispersions of two component 
disks as a function of time using the 1D model of \citet{alexander+2007}. 
This model allow us to consider disks with realistic numbers of stars, but it makes 
a number of simplifying assumptions:
\begin{enumerate}
    \item The disk is axisymmetric.
    \item The disk evolves only via two--body relaxation.
\end{enumerate}
The velocity dispersions ($\sigma$) of the heavy and light stars (indicated by the 
subscripts ``H'' and ``L'') evolve according to the following equations

\begin{align}
    &\frac{d \sigma_H}{d t} = N_H \frac{M_H^2 \log{\Lambda_H}}{a_1 \sigma_H^3(t) t_{\rm orb}}-N_L \frac{M_L M_H \sigma_H(t) \log{\Lambda_{L H}}}{a_2 \sigma_{L H}^4(t) t_{\rm orb}} \left(1-\frac{M_L^2 \sigma_L^2(t)}{M_H^2 \sigma_H^2(t)}\right)\\
    &\frac{d \sigma_L}{d t} = N_L \frac{M_L^2 \log {\Lambda_L}}{a_1 \sigma_L^3(t) t_{\rm orb}}+N_H \frac{M_L M_H \sigma_L(t) \log{\Lambda_{L H}}}{a_2 \sigma_{L H}^4(t) t_{\rm orb}} \left(\frac{M_H^2 \sigma_H^2(t)}{M_L^2 \sigma_L^2(t)}-1\right)\\
    &\sigma_{\rm L H} = \frac{\sigma_L+\sigma_H}{2}\\
    &\Lambda_{X}=\frac{\delta r \sigma_X^2}{2 G M_{\rm X}}\\
    &a_1=3.5 a_2=\frac{2.2 r \delta r}{G^2}
\end{align}
where $\delta r$ is radial extent of the disk, $r$ is its characteristic radius, $M_H$ and $M_L$ are the stellar masses, $t_{\rm orb}$ is the characteristic orbital period, and the $\log{\Lambda_X}$ terms are Coulomb logarithms.

\footnotesize{ 
\bibliographystyle{aastex} 
\bibliography{master_hayden} 

\begin{thebibliography}{}
\expandafter\ifx\csname natexlab\endcsname\relax\def\natexlab#1{#1}\fi
\providecommand{\url}[1]{\href{#1}{#1}}

\bibitem[{{Alexander} {et~al.}(2008){Alexander}, {Armitage}, {Cuadra}, \&
  {Begelman}}]{alexander+2008}
{Alexander}, R.~D., {Armitage}, P.~J., {Cuadra}, J., \& {Begelman}, M.~C. 2008,
  \apj, 674, 927

\bibitem[{{Alexander} {et~al.}(2007){Alexander}, {Begelman}, \&
  {Armitage}}]{alexander+2007}
{Alexander}, R.~D., {Begelman}, M.~C., \& {Armitage}, P.~J. 2007, \apj, 654,
  907

\bibitem[{{Alexander}(2017)}]{alexander2017}
{Alexander}, T. 2017, ArXiv e-prints, arXiv:1701.04762

\bibitem[{{Alexander} \& {Hopman}(2009)}]{alexander&hopman2009}
{Alexander}, T., \& {Hopman}, C. 2009, \apj, 697, 1861

\bibitem[{{Arcavi} {et~al.}(2014){Arcavi}, {Gal-Yam}, {Sullivan}, {Pan},
  {Cenko}, {Horesh}, {Ofek}, {De Cia}, {Yan}, {Yang}, {Howell}, {Tal},
  {Kulkarni}, {Tendulkar}, {Tang}, {Xu}, {Sternberg}, {Cohen}, {Bloom},
  {Nugent}, {Kasliwal}, {Perley}, {Quimby}, {Miller}, {Theissen}, \&
  {Laher}}]{arcavi+2014}
{Arcavi}, I., {Gal-Yam}, A., {Sullivan}, M., {et~al.} 2014, \apj, 793, 38

\bibitem[{{Auchettl} {et~al.}(2017){Auchettl}, {Guillochon}, \&
  {Ramirez-Ruiz}}]{auchettl+2017}
{Auchettl}, K., {Guillochon}, J., \& {Ramirez-Ruiz}, E. 2017, \apj, 838, 149

\bibitem[{{Bahcall} \& {Wolf}(1977)}]{bahcall&wolf1977}
{Bahcall}, J.~N., \& {Wolf}, R.~A. 1977, \apj, 216, 883

\bibitem[{{Bartko} {et~al.}(2010)}]{bartko+2010}
{Bartko}, H., {et~al.} 2010, \apj, 708, 834

\bibitem[{{Chornock} {et~al.}(2014){Chornock}, {Berger}, {Gezari}, {Zauderer},
  {Rest}, {Chomiuk}, {Kamble}, {Soderberg}, {Czekala}, {Dittmann}, {Drout},
  {Foley}, {Fong}, {Huber}, {Kirshner}, {Lawrence}, {Lunnan}, {Marion},
  {Narayan}, {Riess}, {Roth}, {Sanders}, {Scolnic}, {Smartt}, {Smith},
  {Stubbs}, {Tonry}, {Burgett}, {Chambers}, {Flewelling}, {Hodapp}, {Kaiser},
  {Magnier}, {Martin}, {Neill}, {Price}, \& {Wainscoat}}]{chornock+2014}
{Chornock}, R., {Berger}, E., {Gezari}, S., {et~al.} 2014, \apj, 780, 44

\bibitem[{{French} {et~al.}(2016){French}, {Arcavi}, \&
  {Zabludoff}}]{french+2016}
{French}, K.~D., {Arcavi}, I., \& {Zabludoff}, A. 2016, \apjl, 818, L21

\bibitem[{{Gair} {et~al.}(2006){Gair}, {Kennefick}, \& {Larson}}]{gair+2006}
{Gair}, J.~R., {Kennefick}, D.~J., \& {Larson}, S.~L. 2006, \apj, 639, 999

\bibitem[{{Gezari} {et~al.}(2012){Gezari}, {Chornock}, {Rest}, {Huber},
  {Forster}, {Berger}, {Challis}, {Neill}, {et~al.}}]{gezari+2012}
{Gezari}, S., {Chornock}, R., {Rest}, A., {et~al.} 2012, \nat, 485, 217

\bibitem[{Golightly {et~al.}(2019)Golightly, Coughlin, \&
  Nixon}]{golightly+2019}
Golightly, E. C.~A., Coughlin, E.~R., \& Nixon, C.~J. 2019, \apj, 872, 163.
\newblock \url{https://ui.adsabs.harvard.edu/abs/2019ApJ...872..163G}

\bibitem[{{Greene} {et~al.}(2019){Greene}, {Strader}, \& {Ho}}]{greene+2019}
{Greene}, J.~E., {Strader}, J., \& {Ho}, L.~C. 2019, arXiv e-prints,
  arXiv:1911.09678

\bibitem[{{Holoien} {et~al.}(2016){Holoien}, {Kochanek}, {Prieto}, {Grupe},
  {Chen}, {Godoy-Rivera}, {Stanek}, {Shappee}, {Dong}, {Brown}, {Basu},
  {Beacom}, {Bersier}, {Brimacombe}, {Carlson}, {Falco}, {Johnston}, {Madore},
  {Pojmanski}, \& {Seibert}}]{holoien+2016}
{Holoien}, T.~W.-S., {Kochanek}, C.~S., {Prieto}, J.~L., {et~al.} 2016, \mnras,
  463, 3813

\bibitem[{{Hopkins} \& {Quataert}(2010)}]{hopkins&quataert2010}
{Hopkins}, P.~F., \& {Quataert}, E. 2010, \mnras, 407, 1529

\bibitem[{{Kochanek}(2016)}]{kochanek2016}
{Kochanek}, C.~S. 2016, \mnras, 461, 371

\bibitem[{{Kroupa}(2001)}]{kroupa2001}
{Kroupa}, P. 2001, \mnras, 322, 231

\bibitem[{{Kroupa} {et~al.}(2013){Kroupa}, {Weidner}, {Pflamm-Altenburg},
  {Thies}, {Dabringhausen}, {Marks}, \& {Maschberger}}]{kroupa+2013}
{Kroupa}, P., {Weidner}, C., {Pflamm-Altenburg}, J., {et~al.} 2013, {The
  Stellar and Sub-Stellar Initial Mass Function of Simple and Composite
  Populations}, ed. T.~D. {Oswalt} \& G.~{Gilmore}, Vol.~5, 115

\bibitem[{Lauer {et~al.}(1993)Lauer, Faber, Groth, Shaya, Campbell, Code,
  Currie, Baum, Ewald, Hester, Holtzman, Kristian, Light, Ligynds, O'Neil, \&
  Westphal}]{lauer+1993}
Lauer, T.~R., Faber, S.~M., Groth, E.~J., {et~al.} 1993, \aj, 106, 1436.
\newblock \url{https://ui.adsabs.harvard.edu/abs/1993AJ....106.1436L}

\bibitem[{{Levin}(2007)}]{levin2007}
{Levin}, Y. 2007, \mnras, 374, 515

\bibitem[{{Levin} \& {Beloborodov}(2003)}]{levin&beloborodov03}
{Levin}, Y., \& {Beloborodov}, A.~M. 2003, \apjl, 590, L33

\bibitem[{{Lockhart} {et~al.}(2018){Lockhart}, {Lu}, {Peiris}, {Rich},
  {Bouchez}, \& {Ghez}}]{lockhart+2018}
{Lockhart}, K.~E., {Lu}, J.~R., {Peiris}, H.~V., {et~al.} 2018, \apj, 854, 121

\bibitem[{{Lu} {et~al.}(2013){Lu}, {Do}, {Ghez}, {Morris}, {Yelda}, \&
  {Matthews}}]{lu+2013}
{Lu}, J.~R., {Do}, T., {Ghez}, A.~M., {et~al.} 2013, \apj, 764, 155

\bibitem[{{MacLeod} {et~al.}(2012){MacLeod}, {Guillochon}, \&
  {Ramirez-Ruiz}}]{macleod+2012}
{MacLeod}, M., {Guillochon}, J., \& {Ramirez-Ruiz}, E. 2012, \apj, 757, 134

\bibitem[{{Madigan} {et~al.}(2018){Madigan}, {Halle}, {Moody}, {McCourt},
  {Nixon}, \& {Wernke}}]{madigan+2018}
{Madigan}, A.-M., {Halle}, A., {Moody}, M., {et~al.} 2018, \apj, 853, 141

\bibitem[{{Magorrian} {et~al.}(1998){Magorrian}, {Tremaine}, {Richstone},
  {Bender}, {Bower}, {Dressler}, {Faber}, {Gebhardt}, {Green}, {Grillmair},
  {Kormendy}, \& {Lauer}}]{magorrian+1998}
{Magorrian}, J., {Tremaine}, S., {Richstone}, D., {et~al.} 1998, \aj, 115, 2285

\bibitem[{{Merritt}(2013)}]{merritt2013}
{Merritt}, D. 2013, {Dynamics and Evolution of Galactic Nuclei}

\bibitem[{Mikhaloff \& Perets(2017)}]{mikhaloff&perets2017}
Mikhaloff, D.~N., \& Perets, H.~B. 2017, \mnras, 465, 281.
\newblock \url{https://ui.adsabs.harvard.edu/abs/2017MNRAS.465..281M}

\bibitem[{{Miller} \& {Scalo}(1979)}]{miller&scalo1979}
{Miller}, G.~E., \& {Scalo}, J.~M. 1979, \apjs, 41, 513

\bibitem[{{Milosavljevi{\'c}} \& {Loeb}(2004)}]{milosavljevic&loeb2004}
{Milosavljevi{\'c}}, M., \& {Loeb}, A. 2004, \apjl, 604, L45

\bibitem[{{Paumard} {et~al.}(2006){Paumard}, {Genzel}, {Martins}, {Nayakshin},
  {Beloborodov}, {Levin}, {Trippe}, {Eisenhauer}, {Ott}, {Gillessen}, {Abuter},
  {Cuadra}, {Alexander}, \& {Sternberg}}]{paumard+2006}
{Paumard}, T., {Genzel}, R., {Martins}, F., {et~al.} 2006, \apj, 643, 1011

\bibitem[{{Rauch} \& {Tremaine}(1996)}]{rauch&tremaine1996}
{Rauch}, K.~P., \& {Tremaine}, S. 1996, \na, 1, 149

\bibitem[{{Rees}(1988)}]{rees1988}
{Rees}, M.~J. 1988, \nat, 333, 523

\bibitem[{{Rein} \& {Liu}(2012)}]{rein&liu2012}
{Rein}, H., \& {Liu}, S.~F. 2012, \aap, 537, A128

\bibitem[{{Rein} \& {Spiegel}(2015)}]{rein&spiegel2015}
{Rein}, H., \& {Spiegel}, D.~S. 2015, \mnras, 446, 1424

\bibitem[{{Salpeter}(1955)}]{salpeter1955}
{Salpeter}, E.~E. 1955, \apj, 121, 161

\bibitem[{{Spitzer}(1987)}]{spitzer1987}
{Spitzer}, L. 1987, {Dynamical evolution of globular clusters}

\bibitem[{{Stewart} \& {Ida}(2000)}]{ida&stewart2000}
{Stewart}, G.~R., \& {Ida}, S. 2000, \icarus, 143, 28

\bibitem[{{Stone} \& {Metzger}(2016)}]{stone&metzger2016}
{Stone}, N.~C., \& {Metzger}, B.~D. 2016, \mnras, 455, 859

\bibitem[{{Sz{\"o}lgy{\'e}n} \& {Kocsis}(2018)}]{szoelgyen&kocsis2018}
{Sz{\"o}lgy{\'e}n}, {\'A}., \& {Kocsis}, B. 2018, ArXiv e-prints,
  arXiv:1803.07090

\bibitem[{{Tremaine}(1995)}]{tremaine1995}
{Tremaine}, S. 1995, \aj, 110, 628

\bibitem[{{van Velzen} {et~al.}(2011){van Velzen}, {Farrar}, {Gezari},
  {Morrell}, {Zaritsky}, {{\"O}stman}, {Smith}, {Gelfand}, \&
  {Drake}}]{van-velzen+2011}
{van Velzen}, S., {Farrar}, G.~R., {Gezari}, S., {et~al.} 2011, \apj, 741, 73

\bibitem[{{Vasiliev}(2017)}]{vasiliev2017}
{Vasiliev}, E. 2017, \apj, 848, 10

\bibitem[{{Wang} \& {Merritt}(2004)}]{wang&merritt2004}
{Wang}, J., \& {Merritt}, D. 2004, \apj, 600, 149

\bibitem[{{Wernke} \& {Madigan}(2019)}]{wernke&madigan2019}
{Wernke}, H.~N., \& {Madigan}, A.-M. 2019, \apj, 880, 42

\bibitem[{{Wevers} {et~al.}(2019){Wevers}, {Stone}, {van Velzen}, {Jonker},
  {Hung}, {Auchettl}, {Gezari}, {Onori}, {Mata S{\'a}nchez},
  {Kostrzewa-Rutkowska}, \& {Casares}}]{wevers+2019}
{Wevers}, T., {Stone}, N.~C., {van Velzen}, S., {et~al.} 2019, \mnras, 487,
  4136

\end{thebibliography}
}



\end{document}